\documentclass[letterpaper]{article} 
\usepackage{aaai2026}  
\usepackage{times}  
\usepackage{helvet}  
\usepackage{courier}  
\usepackage[hyphens]{url}  
\usepackage{graphicx} 
\usepackage{natbib}  
\usepackage{caption} 
\usepackage{algorithm}
\usepackage{algorithmic}

\usepackage{newfloat}
\usepackage{listings}
\DeclareCaptionStyle{ruled}{labelfont=normalfont,labelsep=colon,strut=off} 
\floatstyle{ruled}
\newfloat{listing}{tb}{lst}{}
\floatname{listing}{Listing}
\usepackage{bm}
\usepackage{amsmath}
\usepackage{amssymb}
\usepackage{diagbox}
\usepackage{subcaption}
\usepackage{booktabs}
\usepackage{graphicx}
\usepackage{multirow}
\usepackage{makecell}

\title{QSIM: Mitigating Overestimation in Multi-Agent Reinforcement Learning\\ via Action Similarity Weighted Q-Learning}

\author {
    Yuanjun Li\textsuperscript{\rm 1},
    Bin Zhang\textsuperscript{\rm 2},
    Hao Chen\textsuperscript{\rm 2},
    Zhouyang Jiang\textsuperscript{\rm 1},
    Dapeng Li\textsuperscript{\rm 3,\rm 4},
    Zhiwei Xu\textsuperscript{\rm 1}\thanks{Corresponding author}
}
\affiliations {
    \textsuperscript{\rm 1}Shandong University\\
    \textsuperscript{\rm 2}The Key Laboratory of Cognition and Decision Intelligence for Complex Systems,\\
    Institute of Automation, Chinese Academy of Sciences\\
    \textsuperscript{\rm 3}Li Auto Inc.\\
    \textsuperscript{\rm 4}Tsinghua University\\
    liyuanjun@mail.sdu.edu.cn, zhiwei\_xu@sdu.edu.cn
}

\usepackage{bibentry}

\begin{document}

\maketitle

\begin{abstract}
\vspace{-3pt} 

Value decomposition (VD) methods have achieved remarkable success in cooperative multi-agent reinforcement learning (MARL). 
However, their reliance on the max operator for temporal-difference (TD) target calculation leads to systematic Q-value overestimation. 
This issue is particularly severe in MARL due to the combinatorial explosion of the joint action space, which often results in unstable learning and suboptimal policies. 
To address this problem, we propose \textbf{QSIM}, a similarity weighted Q-learning framework that reconstructs the TD target using action similarity. 
Instead of using the greedy joint action directly, QSIM forms a similarity weighted expectation over a structured near-greedy joint action space. 
This formulation allows the target to integrate Q-values from diverse yet behaviorally related actions while assigning greater influence to those that are more similar to the greedy choice.
By smoothing the target with structurally relevant alternatives, QSIM effectively mitigates overestimation and improves learning stability.
Extensive experiments demonstrate that QSIM can be seamlessly integrated with various VD methods, consistently yielding superior performance and stability compared to the original algorithms.
Furthermore, empirical analysis confirms that QSIM significantly mitigates the systematic value overestimation in MARL.
Code is available at \url{https://github.com/MaoMaoLYJ/pymarl-qsim}.\looseness=-1
\end{abstract}


\vspace{-10pt} 
\section{Introduction}

Cooperative Multi-Agent Reinforcement Learning (MARL) has emerged as a powerful framework for addressing complex group decision-making problems across domains such as robotic swarm control~\cite{robot1,robot2}, autonomous driving coordination~\cite{autodriving1,autodriving2}, and network resource management~\cite{network1,network2}. 
\begin{figure}[ht!]
    \centering
    \includegraphics[width=0.8\linewidth]{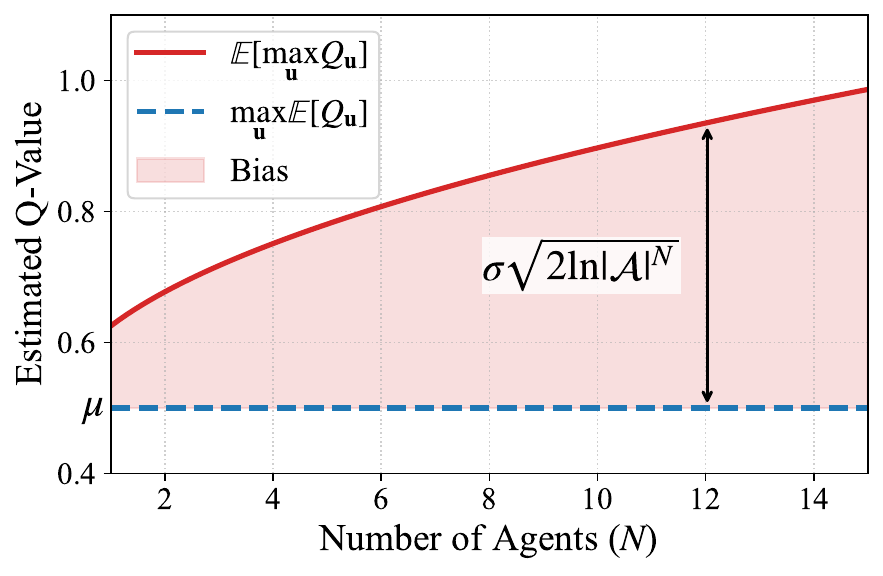}
    \vspace{-10pt} 
    \caption{Theoretical curves showing how the upper bound of overestimation bias scales with the number of agents $N$. Further details are given in Theorem 1.}
    \label{fig:MARL_overestimation}
    \vspace{-20pt} 
\end{figure}
To handle the inherent non-stationarity in multi-agent learning, the Centralized Training with Decentralized Execution (CTDE) paradigm has become the standard approach~\cite{MADDPG,VDN}. 
Within this framework, Value Decomposition (VD) methods~\cite{QMIX} dominate by factorizing the joint action-value function into individual utilities, thereby effectively resolving the credit assignment problem.
Despite their success, these methods suffer from a fundamental limitation: their reliance on the \emph{max operator} when computing the Temporal-Difference (TD) target.
It is well-known in single-agent reinforcement learning that maximizing over noisy value estimates leads to overestimation~\cite{DoubleQlearning}, and this effect becomes even more pronounced in multi-agent settings~\cite{Betterestimation}.
As illustrated in Figure~\ref{fig:MARL_overestimation}, theoretical analysis suggests that the upper bound of the overestimation bias continuously increases with the number of agents $N$, leading to severe learning instability and convergence to suboptimal policies in large-scale scenarios. 
Existing VD methods typically overlook this systemic overestimation error by constructing the TD target based on a single, biased greedy joint action. 
Therefore, addressing this intrinsic overestimation issue is critical.\looseness=-1

The standard practice in VD methods derives the TD target $y$ directly from the Bellman optimality equation, namely $y = r + \gamma \max_{\bm{u}'} Q_{\text{tar}}(s', \bm{u}')$. 
These methods fundamentally rely on the Individual-Global-Max (IGM) principle~\cite{Qtran}, which ensures that the greedy joint action $\bm{u}^*$ can be efficiently obtained through decentralized maximization. 
While this design enables the method to comply with the CTDE paradigm, it remains limited by estimation noise in the learned action-value functions. 
Such noise induces the max operator to preferentially select actions with positively biased estimates~\cite{Overestimation}.
As a result, the calculated target $Q_{\text{tar}}(s', \bm{u}^*)$ becomes biased and systematically overestimates the true expected return. 
A common remedy is to replace the hard maximization with an expectation, following the intuition of Expected SARSA~\cite{ExpectedSARSA}. 
However, implementing such an expectation in MARL is highly impractical: the joint action space grows exponentially with the number of agents, making the expectation both computationally prohibitive and sensitive to errors from rarely sampled actions~\cite{RES}.  
This presents a critical dilemma in that the greedy target remains tractable yet biased, whereas computing a full expectation is theoretically sound but infeasible in practice.

To resolve this dilemma, we propose \textbf{QSIM}, a framework that mitigates overestimation via action similarity weighted Q-learning. 
QSIM reformulates the TD target by leveraging the local structure of the joint action space. 
Instead of relying on a single greedy point estimate or an intractable global expectation, QSIM operates within a constructed \emph{near-greedy} joint action subspace. 
This subspace is composed of single-agent deviations from the greedy policy, characterizes the local topological neighborhood around the greedy action while maintaining linear computational complexity.
Within this subspace, QSIM computes a weighted TD target, where the integration is governed by learned action similarity. 
We posit that actions leading to similar future transitions should share similar values. 
By assigning higher influence to actions that are semantically aligned with the greedy choice, QSIM effectively dampens the noise responsible for overestimation while preserving the optimality of the policy. 

Our main contributions are summarized as follows:
\vspace{-2pt} 

\begin{itemize}
    \item A self-supervised autoencoder is introduced, utilizing a feature encoder to effectively capture the semantic meaning of individual actions. The learned representation provides a reliable similarity metric that quantifies the relationship between joint actions.
    \vspace{-2pt} 
    \item We propose a tractable near-greedy joint action space that scales linearly with the number of agents, upon which a similarity weighted TD target is computed. This design converts the high-variance greedy estimate into a more stable expectation, theoretically proven to constitute a lower bound on the standard greedy estimate, thereby guaranteeing the mitigation of overestimation bias.
    \vspace{-2pt} 
    \item Extensive experiments on multi-agent benchmarks, including SMAC~\cite{SMAC}, SMACv2~\cite{SMACv2}, MPE~\cite{MADDPG} and Matrix Games~\cite{epymarl}, show that integrating QSIM into different value decomposition frameworks consistently improves both performance and training stability over the original methods, while substantially mitigating the overestimation inherent in Q-learning.
\end{itemize}

\vspace{-10pt} 
\section{Related Work}

\subsubsection{Value-based MARL}
Following the CTDE paradigm, VDN~\cite{VDN} established the foundation of value decomposition by approximating the joint Q-value as a sum of individual utilities. 
QMIX~\cite{QMIX} advanced this line of work by introducing a monotonic mixing network that guarantees consistency with the IGM principle.
Subsequent research sought to alleviate the representational limitations of monotonic factorization.
For example, WQMIX~\cite{WQMIX} applies a weighted objective that emphasizes higher-quality joint actions.
QTRAN~\cite{Qtran} reformulates value factorization as a constrained optimization problem.
In addition, QPLEX~\cite{Qplex} leverages a duplex dueling architecture to increase expressiveness within the IGM framework.
Beyond discrete value-based methods, VMIX~\cite{VMIX} extends decomposition to actor-critic frameworks by applying monotonic mixing to a central critic. 
RIIT~\cite{RIIT} further demonstrates that well-optimized monotonic baselines often remain competitive. 
Despite these advances, most prior works concentrate on improving the expressiveness of the value factorization itself. QSIM focuses on a complementary issue by enhancing the quality of the TD target and mitigating the systematic overestimation bias that emerges during training.

\vspace{-5pt} 
\subsubsection{Mitigating Value Overestimation} 
It is well established that applying the max operator to noisy value estimates induces a systematic positive bias~\cite{Overestimation}, and this issue becomes even more pronounced in MARL because of the combinatorial explosion of the joint action space~\cite{Betterestimation}.
Standard single-agent bias mitigation methods like Double Q-learning~\cite{DoubleQ} or Softmax Bellman operators~\cite{SoftmaxBellman} are often insufficient for multi-agent tasks due to the exponential complexity of the action space. 
Consequently, several MARL-specific methods have been developed.
RES~\cite{RES} computes the TD target via a regularized softmax average.  
\(\lambda\)WD QMIX~\cite{WDQMIX} employs weighted double estimators to reduce bias.
And Comix~\cite{Comix} constructs a Sandwich framework that constrains the value function between learnable upper and lower bounds. 
While QSIM adopts the similar near-greedy subspace as RES, it fundamentally differs in how the target is aggregated. 
Whereas RES relies on a softmax operator that can overemphasize overestimated Q-values, QSIM introduces an action similarity mechanism to guide the integration.
This ensures that actions functionally aligned with the greedy policy contribute more to the target, effectively reducing noise from irrelevant actions.

\vspace{-5pt} 
\subsubsection{Action Representation Learning}
A major challenge in MARL is the curse of dimensionality stemming from the exponentially growing joint action space. 
Action representation learning offers a promising solution by learning structured, low-dimensional embeddings that capture relationships among actions. 
In single-agent domains, prior work decompose policies via abstract embeddings~\cite{Actionemb} or model dependencies using hypergraphs~\cite{ActionHypergraph}. 
In multi-agent settings, ROMA~\cite{ROMA} uses latent representations to induce dynamic roles, and RODE~\cite{RODE} clusters actions to build a hierarchical decomposition of the joint space. 
Unlike these methods, which utilize representation learning primarily for policy decomposition or hierarchical control, QSIM employs it for calculating action similarity. 
This similarity then guides the value update, smoothing the TD target with semantically relevant neighboring actions and stabilizing learning.

In summary, QSIM integrates these insights to form a modular framework compatible with diverse VD methods, offering a novel perspective on robust value estimation in high-dimensional multi-agent systems.

\vspace{-2pt} 
\section{Background}

\subsection{MARL Problem Definition}

Cooperative MARL is typically formalized as a Decentralized Partially Observable Markov Decision Process (Dec-POMDP)~\cite{DecPOMDP}, defined by the tuple \( \langle \mathcal{N}, \mathcal{S}, \mathcal{U}, \mathcal{T}, \mathcal{R}, \mathcal{O}, \gamma \rangle \). \(\mathcal{N}\) is a finite set of \(N\) agents, and \(\mathcal{S}\) is the global state space. Each agent \(i \in \mathcal{N}\) draws actions from an individual action space \(\mathcal{A}_i\). At each timestep \(t\), the environment is in state \(s^t \in \mathcal{S}\), and each agent receives a local observation \(o_i^t \in \mathcal{O}_i\). Based on its action-observation history \(\tau_i^t\), each agent \(i\) selects an action \(a_i^t \in \mathcal{A}_i\), forming a joint action \(\bm{u}^t = (a_1^t, \dots, a_N^t) \in \mathcal{U}\), where \(\mathcal{U} = \prod_{i \in \mathcal{N}} \mathcal{A}_i\) represents the joint action space. The system then transitions to the next state \(s^{t+1}\) according to the state transition function \(\mathcal{T}(s^{t+1} | s^t, \bm{u}^t)\) and yields a shared reward \(r^t = \mathcal{R}(s^t, \bm{u}^t)\). The objective is to discover a joint policy \(\boldsymbol{\pi}\) that maximizes the expected discounted return \(J(\boldsymbol{\pi}) = \mathbb{E}_{\pi} [ \sum_{t=0}^{\infty} \gamma^t r^t ]\), where \(\gamma \in (0,1]\) is the discount factor.

This objective is approached by learning the optimal joint action-value function \(Q_{\text{tot}}^*\). For a policy \(\boldsymbol{\pi}\), the joint action-value function \(Q_{\text{tot}}^{\boldsymbol{\pi}}\) is defined as the expected return following the joint action \(\bm{u}^t\) conditioned on the joint history \(\bm{\tau}^t\):
\begin{equation}
    Q_{\text{tot}}^{\boldsymbol{\pi}}(\bm{\tau}^t, \bm{u}^t) = \mathbb{E}_{\boldsymbol{\pi}} \left[ \sum_{k=t}^{\infty} \gamma^{k-t} r^k \mid \bm{\tau}^t, \bm{u}^t \right].
    \label{eq:Qtot}
\end{equation}
The joint history \(\bm{\tau}^t = (\tau_1^t, \dots, \tau_N^t)\) comprises the individual histories \(\tau_i^t = (o_i^1, a_i^1, \dots, o^{t-1}_i, a^{t-1}_i, o^t_i)\), encapsulating all information available to agent \(i\) at timestep \(t\).

\vspace{-3pt} 
\subsection{Centralized Training with Decentralized Execution}

The CTDE framework~\cite{MADDPG} effectively blends the strengths of centralized and decentralized paradigms. This approach trains decentralized agent policies in a centralized fashion, allowing unrestricted access to global information during the training phase. While CTDE algorithms have proven successful in many multi-agent problems~\cite{MAVEN,COCOM}, they face scalability challenges as the joint action-observation space expands exponentially with the number of agents, potentially leading to inefficient learning in large-scale scenarios.

Value decomposition has emerged as a key approach within the CTDE framework. It employs a mixing network to compose individual utility functions \(Q_i(\tau_i, a_i)\) into a factored joint action-value function \(Q_{\text{tot}}(\bm{\tau}, \bm{u})\). Most value decomposition methods are built upon the IGM assumption, which states that the global optimum of the factored value function coincides with the set of local optima for the individual utility functions:
\begin{equation}
\arg \max_{\bm{u}} Q_{\text{tot}}(\bm{\tau}, \bm{u}) = 
\begin{pmatrix} 
\arg \max_{a_1} Q_1(\tau_1, a_1) \\[2pt]
\arg \max_{a_2} Q_2(\tau_2, a_2) \\[2pt]
\vdots \\[2pt]
\arg \max_{a_N} Q_N(\tau_N, a_N)
\end{pmatrix}.
\end{equation}
This property is essential for enabling efficient decentralized execution.
The parameters \(\theta\) of the networks are trained by minimizing the TD-error loss over episodes sampled from a replay buffer:
\begin{equation}
\label{eq:td_loss}
\mathcal{L}(\theta) = \mathbb{E} \left[ \left( y - Q_{\text{tot}}(\bm{\tau}, \bm{u}; \theta) \right)^2 \right],
\end{equation}
where the TD target \(y\) is defined as:
\begin{equation}
\label{eq:td_target}
y = r + \gamma \max_{\bm{u'}} Q_{\text{tar}}(\bm{\tau'}, \bm{u'}; \theta^-).
\end{equation}
\(Q_{\text{tot}}(\bm{\tau}, \bm{u}; \theta)\) denotes the value produced by the main network, and the target value \(Q_{\text{tar}}(\bm{\tau'}, \bm{u'}; \theta^-)\) is generated by a target network with frozen parameters \(\theta^-\). The target parameters are periodically updated from the main network to improve training stability~\cite{DQN}.

\begin{figure*}[t]
  \centering
  \includegraphics[width=\textwidth]{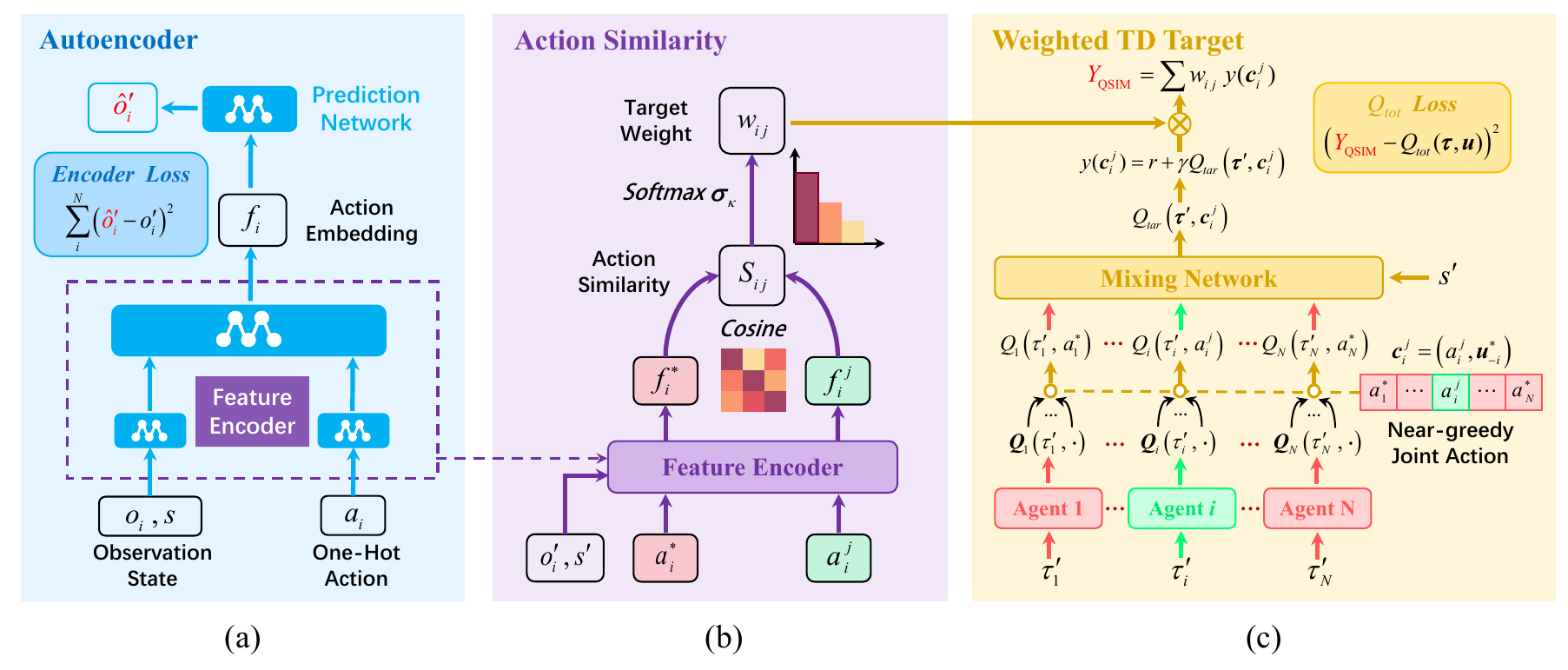}
    \caption{QSIM framework. \textbf{(a) Autoencoder:} Self-supervised learning of action representations. \textbf{(b) Action Similarity:} Computing cosine similarity between deviating action $a^j_i$ and greedy action $a^*_i$ to derive softmax-normalized weights. \textbf{(c) Weighted TD Target:} Constructing near-greedy joint actions $\bm{c}^j_i$ and aggregating their TD target into the final weighted TD target $Y_{\text{QSIM}}$.\looseness=-1}
  \label{fig:qsim_framework}
\end{figure*}

\vspace{-5pt} 
\subsection{Overestimation in Q-learning}
\label{sec:overestimation}
\vspace{-3pt} 

Maximization bias arises from applying the max operator to noisy value estimates. Let \(\hat{Q}(s,a) = Q^*(s,a) + \epsilon_a\) be the learned approximation corrupted by zero-mean noise \(\epsilon_a\). The standard TD target \(y = r + \gamma \max_{a'} \hat{Q}(s', a')\) introduces a positive bias due to Jensen's inequality:
\begin{equation}
    \begin{split}
        \mathbb{E}[\max_{a'} \hat{Q}(s', a')] 
        &= \mathbb{E}[\max_{a'} (Q^*(s', a') + \epsilon_{a'})] \\
        &\ge \max_{a'} \mathbb{E}[Q^*(s', a') + \epsilon_{a'}] \\
        &= \max_{a'} Q^*(s', a').
    \end{split}
    \label{eq:jensen}
\end{equation}
This inequality demonstrates that the target systematically overestimates the true maximum action value.

This bias becomes substantially more severe in MARL due to the combinatorial explosion of the joint action space. 
Consider a system with $N$ agents, each possessing an action space of size $|\mathcal{A}|$, resulting in a joint action space size of $|\mathcal{U}| = |\mathcal{A}|^N$. 
To quantify the impact of this exponential expansion on value estimation, we analyze the bias under the standard assumption that estimation errors follow a Gaussian distribution~\cite{Gaussian_Qvalue}.

\noindent\textbf{Theorem 1.} \textit{Let the estimated joint Q-value be modeled as $\hat{Q}(\bm{u}) = Q^*(\bm{u}) + \epsilon_{\bm{u}}$ for any joint action $\bm{u} \in \mathcal{U}$, where $\epsilon_{\bm{u}} \sim \mathcal{N}(0, \sigma^2)$ denotes independent zero-mean Gaussian noise. The maximization bias is upper bounded as follows:}
\begin{equation*}
    \left(\mathbb{E}\left[\max_{\bm{u}} \hat{Q}(\bm{u})\right] - \max_{\bm{u}} Q^*(\bm{u})\right) \le \sigma \sqrt{2 \ln |\mathcal{A}|^N}.
    \label{eq:bias_theorem}
\end{equation*}

A detailed proof is provided in Appendix~\ref{app:bias_proof}. 
The theorem establishes that the upper bound of the overestimation bias grows proportionally to $\sqrt{N}$ (since $\sqrt{\ln |\mathcal{A}|^N} = \sqrt{N \ln |\mathcal{A}|}$). 
Figure~\ref{fig:MARL_overestimation} illustrates this theoretical trend. 
As the number of agents increases, the potential gap between the expected greedy estimate and the true value expands accordingly. 
This confirms that the exponential growth of the multi-agent action space significantly amplifies overestimation, ultimately weakening the decentralized policies trained under these biased central estimates.


\section{QSIM Framework}

To mitigate the maximization bias in VD methods, we introduce the \textbf{QSIM} framework. It replaces the high-variance greedy TD target with a robust, similarity weighted expectation over a tractable near-greedy joint action space. By transitioning from a biased point estimate to a smoothed expectation, QSIM dampens the noise responsible for overestimation, promoting stable and accurate value updates. The framework consists of three components: (1) a self-supervised autoencoder learning functional action representations; (2) an action similarity mechanism for deriving integration weights from learned action embeddings; (3) a similarity weighted aggregation scheme for the final TD target. The overall architecture is illustrated in Figure~\ref{fig:qsim_framework}.

\subsection{Learning Action Representations}
\label{sec:action_representations}

To quantify semantic similarity between actions, we learn embeddings that describe how each action influences future observations.
The key idea is that the meaning of an action is reflected in the state changes it causes.
We employ a self-supervised autoencoder parameterized by $\phi$, utilizing the next observation as a training signal. This module is visualized in Figure~\ref{fig:qsim_framework}(a).

Formally, the architecture consists of a feature encoder \(E_{\phi}\) and a predictive network \(P_{\phi}\). 
For each agent \(i\), the encoder \(E_{\phi}\) processes the local observation \(o_i\), the global state \(s\), and the chosen action \(a_i\). 
As implemented, \(o_i\) and \(s\) are encoded in parallel with \(a_i\) before being merged to produce a low-dimensional state-action embedding \(f_i\) formulated as:
\begin{equation}
    f_i = E_{\phi}(o_i, s, a_i).
    \label{eq:encoder}
\end{equation}
To ensure these embeddings capture global transition dynamics, the embeddings from all agents \(\bm{f} = (f_1, \dots, f_N)\) are concatenated and passed to the prediction network \(P_{\phi}\). This network reconstructs the next joint local observation \(\hat{\bm{o}}' = (\hat{o}'_1, \dots, \hat{o}'_N)\):
\begin{equation}
    \hat{\bm{o}}' = P_{\phi}(\bm{f}).
    \label{eq:decoder}
\end{equation}
The autoencoder is optimized by minimizing the Mean Squared Error (MSE) between the predicted and actual next observations:
\begin{equation}
    \mathcal{L}_{\text{AE}}(\phi) = \mathbb{E} \left[ \sum_{i=1}^{N} \left( \hat{o}'_i - o'_i \right)^2 \right].
    \label{eq:ae_loss}
\end{equation}
Minimizing this loss encourages each \(f_i\) to encode predictive information about the transition, yielding a reliable metric for action similarity. Empirically, this objective converges rapidly, as illustrated in Appendix~\ref{app:ae_convergence}, enabling QSIM to utilize stable and informative similarity metrics from the early stages of training.

\subsection{Near-greedy Joint Action Space}
\label{sec:suboptimal_space}

Traditional VD methods rely on a single greedy TD target. While computing an expectation over the full space \(\mathcal{U}\) is theoretically appealing, it is computationally intractable and highly sensitive to estimation noise due to sparse visitation. 
To address these issues, we construct a tractable near-greedy joint action space \(\mathcal{C}\) anchored on the greedy joint action, focusing the expectation on the most plausible region. 
Following the Double Q-learning, we first identify the greedy action \(\bm{u}^*\) using the main network, which is efficiently computed via local greedy selections under the IGM principle. 
Given \(\bm{\tau}'\), the greedy joint action \(\bm{u}^*\) is defined as:
\begin{equation}
    \bm{u}^* = \operatorname*{arg\,max}_{\bm{u} \in \mathcal{U}} \ Q_{\text{tot}}(\bm{\tau}', \bm{u}; \theta).
    \label{eq:greedy_action_selection}
\end{equation}

The near-greedy joint action space \(\mathcal{C}\) is then formed by considering all possible single-agent deviating actions from \(\bm{u}^*\). 
Specifically, it is the union of joint actions where exactly one agent \(i\) deviates to an available action \(a_i \in \mathcal{A}_i\), while all other agents $\mathcal{N}_{-i}$ maintain their greedy actions \(\bm{u}^*_{-i}\):
\begin{equation}
    \mathcal{C} = \bigcup_{i=1}^{N} \{ (a_i, \bm{u}^*_{-i}) \mid a_i \in \mathcal{A}_i \}.
    \label{eq:suboptimal_space_def}
\end{equation}
The size of the near-greedy action space grows linearly with the number of agents, $|\mathcal{C}| = N \times |\mathcal{A}|$, in sharp contrast to the exponential growth of the full joint action space $|\mathcal{U}| = |\mathcal{A}|^N$. 
Additional discussion of the near-greedy space $\mathcal{C}$ is provided in Appendix~\ref{app:action_space}.

Finally, each near-greedy joint action \(\bm{c} \in \mathcal{C}\) is evaluated to generate a TD target candidate \(y(\bm{c})\):
\begin{equation}
    y(\bm{c}) = r + \gamma Q_{\text{tar}}(\bm{\tau}', \bm{c}; \theta^-),
    \label{eq:individual_td_target}
\end{equation}
where \(Q_{\text{tar}}(\bm{\tau}', \bm{c}; \theta^-)\) denotes the global joint action-value predicted by the target network. 
This process yields a diverse set of target candidates, which are subsequently aggregated as detailed in the next section.

\subsection{Weighted Q-learning with Action Similarity}
\label{sec:weighted_q_learning}

Upon constructing the near-greedy joint action space \(\mathcal{C}\) and generating the set of TD target candidates, the final phase of QSIM is to synthesize these candidates into a single robust learning signal. This is achieved by computing a weighted expectation in which each weight reflects the functional similarity between a near-greedy joint action and the greedy joint action. This weighting scheme ensures that the value update is regularized by diverse actions, while giving higher importance to actions that are most functionally aligned with the greedy choice.

\subsubsection{Similarity Calculation}
First, we leverage the feature encoder \(E_{\phi}\) to quantify the functional semantics of actions in the next timestep. 
For each agent \(i\), let \(a^*_{i}\) denote its component in the greedy joint action \(\bm{u}^*\), and \(a^j_i \in \mathcal{A}_i\) denote any available action. 
Conditioned on the next global state \(s'\) and local observation \(o'_i\), we compute the greedy action embedding \(f^*_{i}\) and the deviating action embedding \(f^j_i\):
\begin{equation}
    \label{eq:embedding}
    \begin{split}
        f^*_{i} &= E_{\phi}(o'_{i}, s', a^*_{i}), \\
        f^j_{i} &= E_{\phi}(o'_{i}, s', a^j_{i}).
    \end{split}
\end{equation}
The similarity score \(S_{ij}\) is then computed using cosine similarity between these embeddings:
\begin{equation}
    S_{ij} = \text{Cosine}(f^*_{i}, f^j_{i}) = \frac{f^*_{i} \cdot f^j_{i}}{\| f^*_{i} \| \| f^j_{i} \|}.
    \label{eq:similarity}
\end{equation}

To extend the learned individual action similarity metrics to the joint action space, the specific composition of the near-greedy space $\mathcal{C}$ is first formalized. 
A near-greedy joint action $\bm{c}^j_{i} \in \mathcal{C}$ denotes the configuration where agent $i$ selects an deviating action $a^j_{i} \in \mathcal{A}_i$, while all other agents maintain their greedy actions $\bm{u}^*_{-i}$:
\begin{equation}
    \bm{c}^j_{i} = (a^j_{i}, \bm{u}^*_{-i}).
\end{equation}
Given this, the relationship between joint action similarity and individual action similarity is established as follows:

\noindent\textbf{Definition 1} (Action Similarity). \textit{The functional similarity between a near-greedy joint action \(\bm{c}^j_{i}\) and the greedy joint action \(\bm{u}^*\) is defined as the local action similarity of the deviating agent:}
\begin{equation}
    \text{Sim}(\bm{c}^j_{i}, \bm{u}^*) \triangleq \text{Sim}(a^j_{i}, a^*_{i}) = S_{ij}.
\end{equation}
This definition ensures that the similarity score is fully determined by the deviating action of the specific agent modifying its policy, as the behaviors of all other agents remain constant.\looseness=-1

\subsubsection{Similarity Weighted Aggregation}
We transform the raw similarity scores into a normalized probability distribution to derive the integration weights. This is achieved using a softmax function with an inverse temperature parameter \(\kappa \ge 0\):\looseness=-1
\begin{equation}
    w_{ij} = \frac{\exp(\kappa \cdot S_{ij})}{\sum_{k=1}^{|\mathcal{A}_i|} \exp(\kappa \cdot S_{ik})}.
    \label{eq:softmax_weights}
\end{equation}
The hyperparameter \(\kappa\) controls the sharpness of the weight distribution. As \(\kappa \to \infty\), the distribution becomes increasingly peaked, approximating a hard max operator in which the action most similar to the greedy choice dominates. Conversely, as \(\kappa \to 0\), the distribution becomes uniform, yielding a uniform weight expectation over the near-greedy joint action space.

Finally, the QSIM weighted TD target \(Y_{\text{QSIM}}\) is computed as the weighted value of the target candidates over the near-greedy space \(\mathcal{C}\). 
Let \(w_{ij}\) denote the weight associated with the near-greedy joint action \(\bm{c}^j_{i}\). The final target is given by:
\begin{equation}
    Y_{\text{QSIM}} = \sum_{\bm{c}^j_{i} \in \mathcal{C}} w_{ij} \cdot y(\bm{c}^j_{i}),
    \label{eq:final_target}
\end{equation}
where \(y(\bm{c}^j_{i})\) is the candidate target defined in Eq.~\eqref{eq:individual_td_target}. 
This aggregated target \(Y_{\text{QSIM}}\) replaces the standard greedy target in the TD loss function provided in Eq.~\eqref{eq:td_loss}. 
By incorporating value estimates from the neighborhood of the greedy policy, the resulting update provides a principled mechanism for mitigating overestimation bias.

\noindent\textbf{Theorem 2.} \textit{The value estimation component of the QSIM TD target, denoted as \(V_{\text{QSIM}}(s')\), constitutes a lower bound on the standard greedy value estimate \(V_{\text{Greedy}}(s')\). Formally, since \(\sum_{\bm{c} \in \mathcal{C}} w(\bm{c}) = 1\) and \(w(\bm{c}) \ge 0\), it holds that:}
\begin{align*}
    V_{\text{QSIM}}(s') \leq V_{\text{Greedy}}(s').
\end{align*}
This theorem demonstrates that the QSIM operator systematically reduces the value estimate compared to the greedy max operator. A complete proof is provided in Appendix~\ref{app:bias_analysis}.

By leveraging this similarity weighted expectation, QSIM effectively exploits the local structure of the value function around the greedy policy. 
Instead of relying on a single potentially unstable point estimate, it synthesizes a consensus target from functionally similar actions. 
This approach not only provides a strictly lower-variance learning signal to mitigate overestimation but also ensures that the value update remains grounded in the agents' semantic behavior. 
The complete training procedure is summarized in Algorithm~\ref{alg:qsim} as shown in Appendix~\ref{app:pseudocode}.

\begin{figure*}[!htbp]
    \centering
    \includegraphics[width=\textwidth]{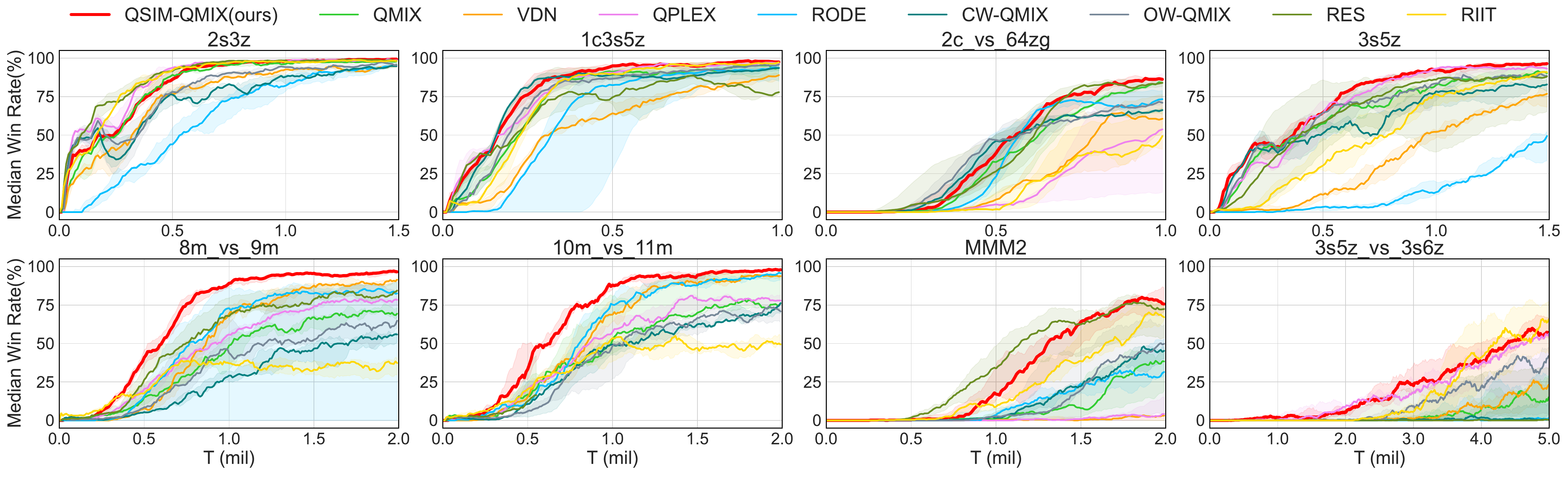}
    \caption{Performance comparison on SMAC maps.}
    \label{fig:smac_results}
\end{figure*}

\begin{table*}[!t]
\centering
\resizebox{\textwidth}{!}{%
\begin{tabular}{c|ccccccccc|c}
\toprule
\multicolumn{1}{c|}{\textbf{Task}} & \textbf{QSIM-QMIX} & \textbf{QMIX} & \textbf{VDN} & \textbf{QPLEX} & \textbf{RODE} & \textbf{CW-QMIX} & \textbf{OW-QMIX} & \textbf{RES} & \textbf{RIIT} & \textbf{Steps} \\
\midrule
\textit{2s3z}           & \textbf{99.4}{\scriptsize \textbf{(0.2)}} & 97.9{\scriptsize (0.7)} & 95.8{\scriptsize (1.4)} & 98.8{\scriptsize (0.6)} & 95.6{\scriptsize (1.6)} & 94.8{\scriptsize (2.6)} & 95.4{\scriptsize (1.3)} & 98.3{\scriptsize (0.5)} & 98.1{\scriptsize (0.9)} & 1.5e6 \\
\textit{1c3s5z}         & \textbf{97.5}{\scriptsize \textbf{(0.8)}} & 93.8{\scriptsize (4.1)} & 89.0{\scriptsize (2.8)} & 96.5{\scriptsize (1.9)} & 93.8{\scriptsize (13.9)} & 93.5{\scriptsize (3.2)} & 95.8{\scriptsize (1.4)} & 77.9{\scriptsize (28.6)} & 97.3{\scriptsize (3.0)} & 1.0e6 \\
\textit{2c\_vs\_64zg}   & \textbf{86.3}{\scriptsize \textbf{(2.0)}} & 84.0{\scriptsize (17.8)} & 60.6{\scriptsize (19.0)} & 53.8{\scriptsize (36.0)} & 73.3{\scriptsize (27.5)} & 66.3{\scriptsize (2.2)} & 71.0{\scriptsize (15.6)} & 84.0{\scriptsize (3.4)} & 49.4{\scriptsize (15.4)} & 1.0e6 \\
\textit{3s5z}           & \textbf{96.5}{\scriptsize \textbf{(1.1)}} & 90.2{\scriptsize (2.6)} & 76.9{\scriptsize (7.6)} & 93.1{\scriptsize (2.2)} & 49.4{\scriptsize (15.2)} & 82.9{\scriptsize (5.0)} & 88.3{\scriptsize (2.4)} & 88.1{\scriptsize (28.0)} & 90.8{\scriptsize (3.6)} & 1.5e6 \\
\textit{8m\_vs\_9m}    & \textbf{96.6}{\scriptsize (2.2)} & 69.4{\scriptsize (8.4)} & 91.5{\scriptsize \textbf{(2.0)}} & 78.3{\scriptsize (3.3)} & 82.5{\scriptsize (43.4)} & 55.8{\scriptsize (8.8)} & 64.8{\scriptsize (17.0)} & 84.4{\scriptsize (6.2)} & 37.3{\scriptsize (9.3)} & 2.0e6 \\
\textit{10m\_vs\_11m}  & \textbf{97.9}{\scriptsize \textbf{(1.4)}} & 76.0{\scriptsize (9.9)} & 93.8{\scriptsize (2.9)} & 77.9{\scriptsize (9.3)} & 95.6{\scriptsize (2.3)} & 76.3{\scriptsize (8.6)} & 70.8{\scriptsize (7.6)} & 16.5{\scriptsize (24.7)} & 49.4{\scriptsize (7.8)} & 2.0e6 \\
\textit{MMM2}           & \textbf{75.6}{\scriptsize \textbf{(8.4)}} & 38.3{\scriptsize (27.9)} & 3.1{\scriptsize (19.8)} & 4.2{\scriptsize (20.8)} & 31.5{\scriptsize (24.2)} & 45.2{\scriptsize (20.5)} & 49.8{\scriptsize (31.3)} & 72.4{\scriptsize (33.4)} & 67.1{\scriptsize (10.1)} & 2.0e6 \\
\textit{3s5z\_vs\_3s6z} & 57.1{\scriptsize \textbf{(8.8)}} & 15.0{\scriptsize (13.9)} & 23.1{\scriptsize (21.8)} & 56.5{\scriptsize (23.8)} & 1.5{\scriptsize (20.7)} & 0.4{\scriptsize (14.9)} & 41.9{\scriptsize (24.0)} & 0.4{\scriptsize (32.1)} & \textbf{65.6}{\scriptsize (22.9)} & 5.0e6 \\
\bottomrule
\end{tabular}%
}
\caption{Performance comparison on the SMAC benchmark, where results are reported as the final median test win rate (\%) with standard deviation over 5 random seeds.}
\label{tab:smac_winrate}
\end{table*}


\section{Experiments}
\label{sec:experiments}

We evaluate the proposed QSIM framework primarily using the QMIX backbone, denoted as \textbf{QSIM-QMIX}. Our empirical analysis addresses five key research questions:
\begin{enumerate}
    \item Performance \& Stability: Does QSIM-QMIX outperform existing VD baselines in both win rate and stability?
    \item Generality: Does integrating QSIM with other VD backbones (VDN, QPLEX) consistently yield gains across diverse environments?
    \item Ablation: How critical are the near-greedy action space and the similarity weighted aggregation scheme?
    \item Overestimation: Does QSIM effectively mitigate the overestimation bias? We analyze the error between estimated $\hat{Q}_{\text{tot}}$ and actual discounted return.
    \item Action Representation Visualization: Do the learned action embeddings capture meaningful action semantics?
\end{enumerate}

\subsubsection{Benchmarks and Baselines} Our empirical evaluation spans four diverse benchmarks: SMAC~\cite{SMAC}, SMACv2~\cite{SMACv2}, MPE~\cite{MADDPG} and Matrix Games~\cite{epymarl}.
Comparative analysis is conducted against a comprehensive suite of baselines, including VDN~\cite{VDN}, QMIX~\cite{QMIX}, WQMIX~\cite{WQMIX}, QPLEX~\cite{Qplex}, RES~\cite{RES}, RODE~\cite{RODE}, and RIIT~\cite{RIIT}.

\subsubsection{Experimental Setup} Experiments are conducted over 5 random seeds. Plots report median performance (solid line) with 25th--75th percentile interquartile ranges (shaded area), where smaller shaded areas indicate higher stability. Additional training details are provided in Appendix~\ref{app:experiment_details}.

\begin{figure*}[ht!]
    \centering
    \includegraphics[width=0.9\linewidth]{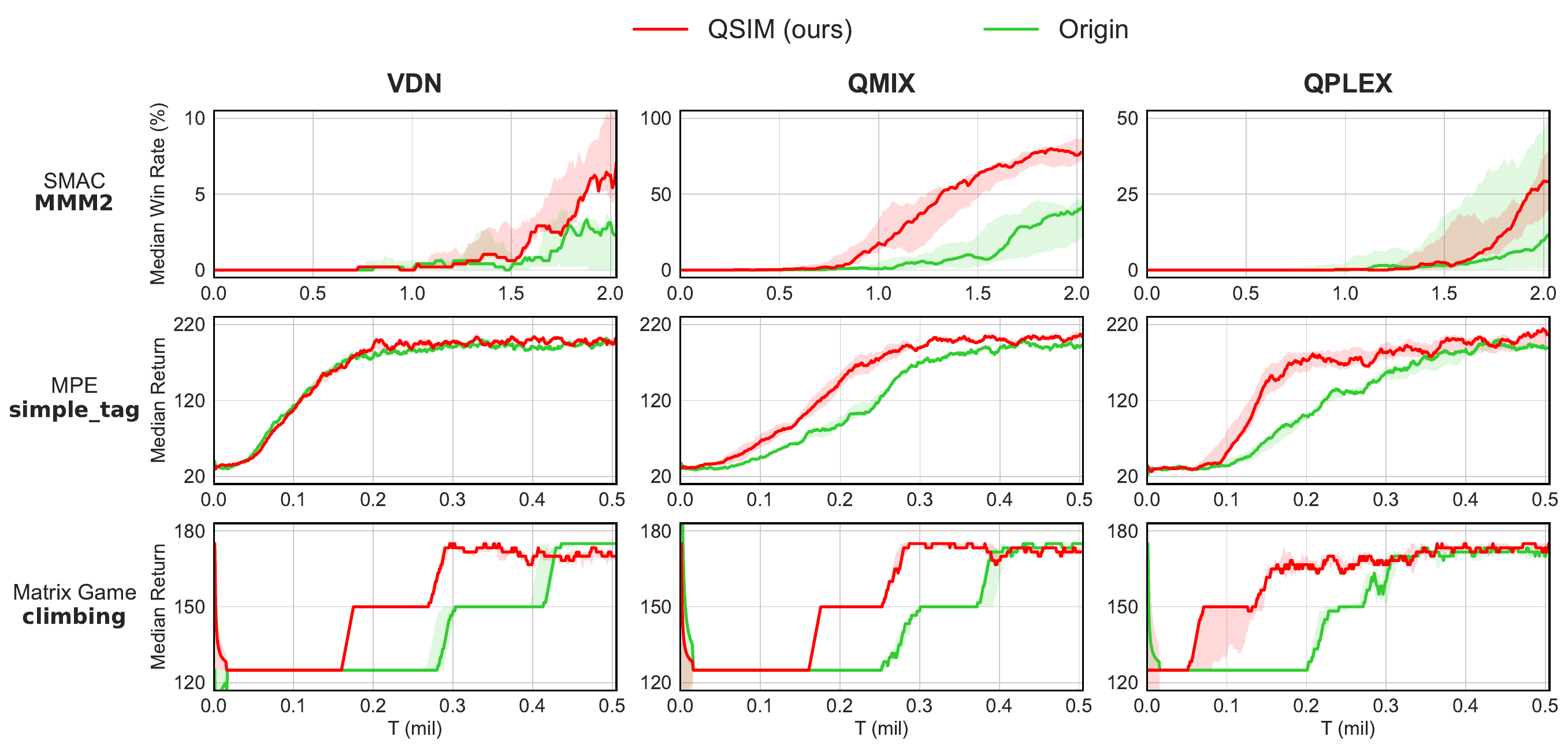}
    \caption{Comparison of QSIM-enhanced variants with their original baselines across different benchmarks.}
    \label{fig:generality_results}
\end{figure*}

\subsection{Performance on SMAC}

We evaluate on SMAC, a benchmark requiring fine-grained micromanagement under partial observability.
The combinatorial explosion in SMAC poses a severe challenge for value estimation. 
To ensure a comprehensive evaluation, we select eight maps spanning a spectrum of difficulty: two easy (\textit{2s3z}, \textit{1c3s5z}), four hard (\textit{2c\_vs\_64zg}, \textit{3s5z}, \textit{8m\_vs\_9m}, \textit{10m\_vs\_11m}), and two super-hard (\textit{MMM2}, \textit{3s5z\_vs\_3s6z}).

Figure~\ref{fig:smac_results} and Table~\ref{tab:smac_winrate} collectively illustrate the superior performance and stability of QSIM-QMIX. 
In high-dimensional scenarios where standard VD methods are prone to instability driven by maximization bias, QSIM-QMIX achieves higher mean win rates with significantly lower standard deviation. 
To further evaluate the robustness and generalization capabilities of QSIM, we extend our experiments to the SMACv2 benchmark~\cite{SMACv2}. SMACv2 introduces significant environmental stochasticity, including randomized unit types and start positions. This dynamic setting requires agents to develop adaptable cooperative strategies capable of handling unseen configurations. Detailed empirical results on SMACv2 are provided in Appendix~\ref{app:smacv2}.\looseness=-1

\subsection{Generality of QSIM}
\label{sec:generality}

A pivotal advantage of QSIM is its modular design, allowing seamless integration into various VD methods. 
To empirically verify this generality, we extend our evaluation beyond QMIX, integrating QSIM with two other prominent VD backbones: VDN and QPLEX. 
We then compare the resulting variants (QSIM-VDN, QSIM-QMIX, QSIM-QPLEX) with their original counterparts across multiple environments. 
Detailed information about the experimental environment is shown in Appendix~\ref{app:env_intro}.

Figure~\ref{fig:generality_results} summarizes the comparative results on three representative scenarios covering different domains. 
Comprehensive results for additional scenarios are provided in Appendix~\ref{app:add_generality}.
Across all tested environments, the QSIM-enhanced variants consistently outperform or match the original baselines.
These findings demonstrate that QSIM is broadly compatible with existing value decomposition frameworks and reliably boosts their performance.
By providing a more robust learning signal, QSIM improves the sample efficiency and stability of arbitrary value decomposition architectures, regardless of the task complexity.

\subsection{Ablation Study}
\label{sec:ablation_similarity}

We conduct an ablation study to isolate the contributions of two core mechanisms, namely the construction of the near-greedy joint action space and the similarity weighted scheme. 
To this end, we introduce QSIM-Mean, a variant that retains the near-greedy space structure but replaces the similarity weight with a uniform weight $w(\bm{c}) = 1 / |\mathcal{C}|$. 
This is equivalent to setting the inverse temperature $\kappa=0$ in Eq.~\eqref{eq:softmax_weights}, thereby assigning the same weight to each near-greedy joint action $\bm{c}$. 
Additional QSIM variants and extended ablation studies are provided in Appendix~\ref{app:add_abalation}.

\begin{figure}[ht!]
    \centering
    \includegraphics[width=\linewidth]{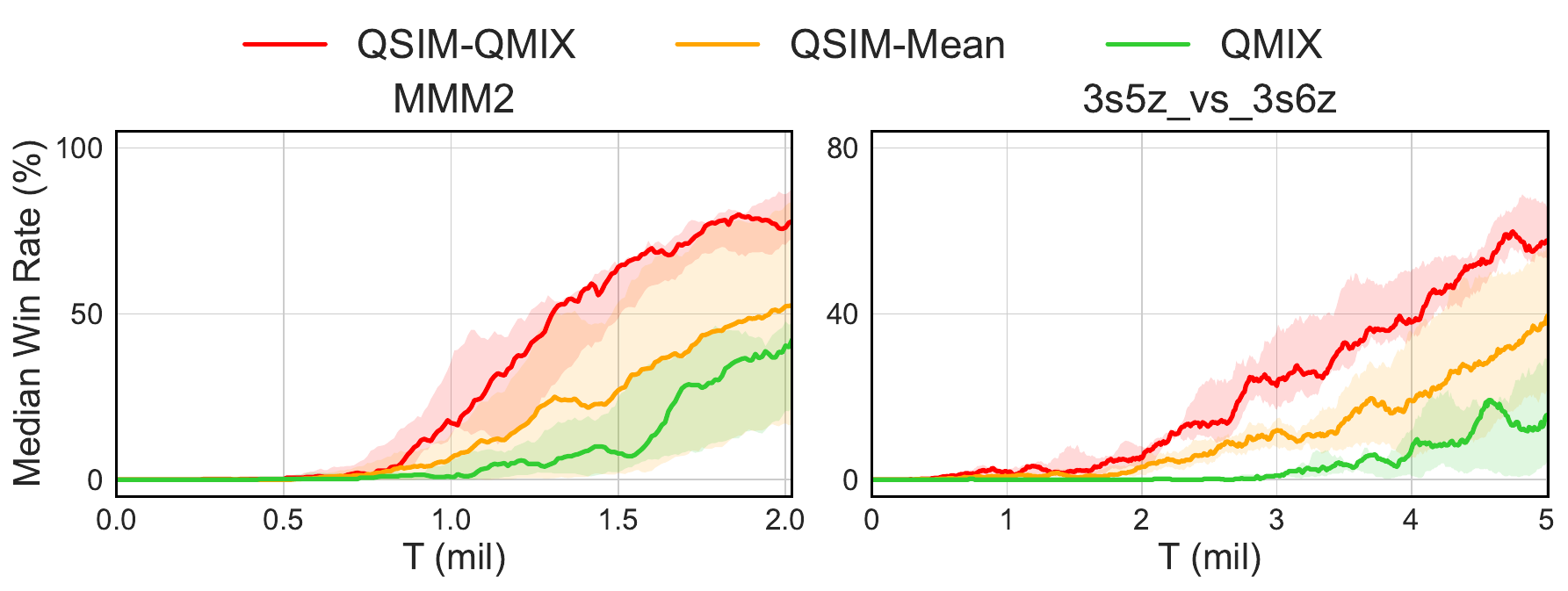}
    \caption{Ablation study comparing the full QSIM-QMIX model against the unweighted QSIM-Mean variant and the original QMIX baseline.}
    \label{fig:QSIM-Mean}
    \vspace{-15pt}
\end{figure}

As shown in Figure~\ref{fig:QSIM-Mean}, the results across SMAC tasks reveal a clear performance ordering in which QSIM-QMIX outperforms QSIM-Mean, and both variants outperform QMIX. This observation provides two important insights.

First, the improvement of QSIM-Mean over QMIX highlights the structural benefit brought by the near-greedy action space.
Since $\mathcal{C}$ consists of single-agent deviations, it captures the local neighborhood of the greedy policy. 
Aggregating over these near-greedy actions provides a more diverse and better-regularized learning signal. 
Compared with the single-point greedy estimate used by QMIX, this smoothed expectation is more robust and helps avoid premature convergence to suboptimal solutions.

Second, the further gains achieved by QSIM-QMIX over QSIM-Mean underscore the importance of similarity weighting.
While simple averaging reduces computational overhead, it naively aggregates all actions, inevitably including irrelevant or counter-productive deviating actions. 
By assigning higher weights to actions functionally similar to the greedy policy, QSIM effectively filters out noisy or implausible deviations. 
This selective aggregation yields a more informative target, enabling faster and more stable convergence.\looseness=-1

\begin{figure*}[ht!]
    \centering
    \includegraphics[width=\linewidth]{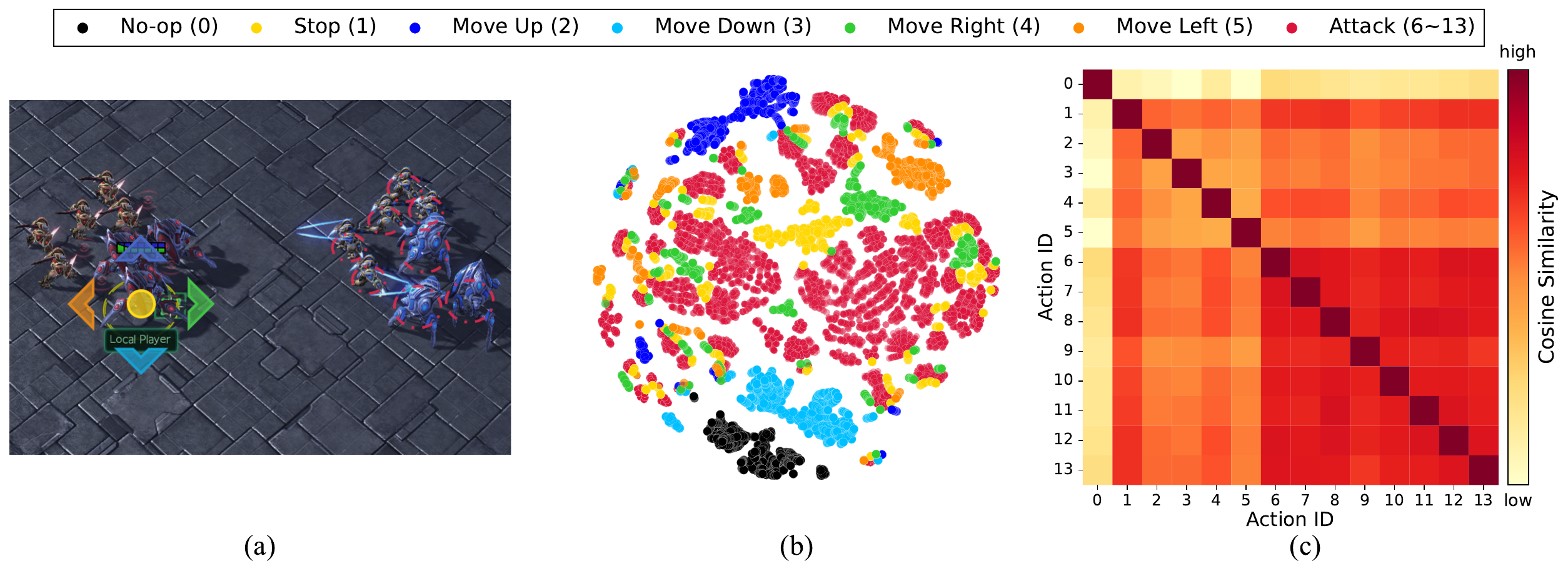}
    \caption{Visualization of learned action embeddings for a Stalker agent in the SMAC \textit{3s5z} scenario.
    (a) The action space of a Stalker agent in the \textit{3s5z} scenario.
    (b) A projection of the learned action embeddings produced via t-SNE.
    (c) Similarity matrix between all actions of the Stalker agent.}
    \label{fig:embedding_viz}
\end{figure*}

\subsection{Mitigation of Q-Value Overestimation}
\label{sec:overestimation_analysis}

To evaluate QSIM's ability to mitigate the systematic Q-value overestimation in VD methods, we analyze the estimation error $\delta_q$, defined as the error between the estimated joint Q-value $\hat{Q}_{\text{tot}}$ and the actual discounted return:
\begin{equation}
    \delta_q = \hat{Q}_{\text{tot}}(\bm{\tau}_t, \bm{u}_t) - \sum_{k=t}^{\infty} \gamma^{k-t} r_k,
    \label{eq:estimation_error}
\end{equation}
where the summation term approximates the actual joint Q-value $Q^{\pi}_{\text{tot}}$ in Eq.~\eqref{eq:Qtot}. 
We compare the resulting estimation error $\delta_q$ of QSIM-QMIX with that of QMIX on several challenging SMAC maps by examining how the error evolves over training steps.

\begin{figure}[ht!]
    \vspace{-10pt}
    \centering
    \includegraphics[width=\linewidth]{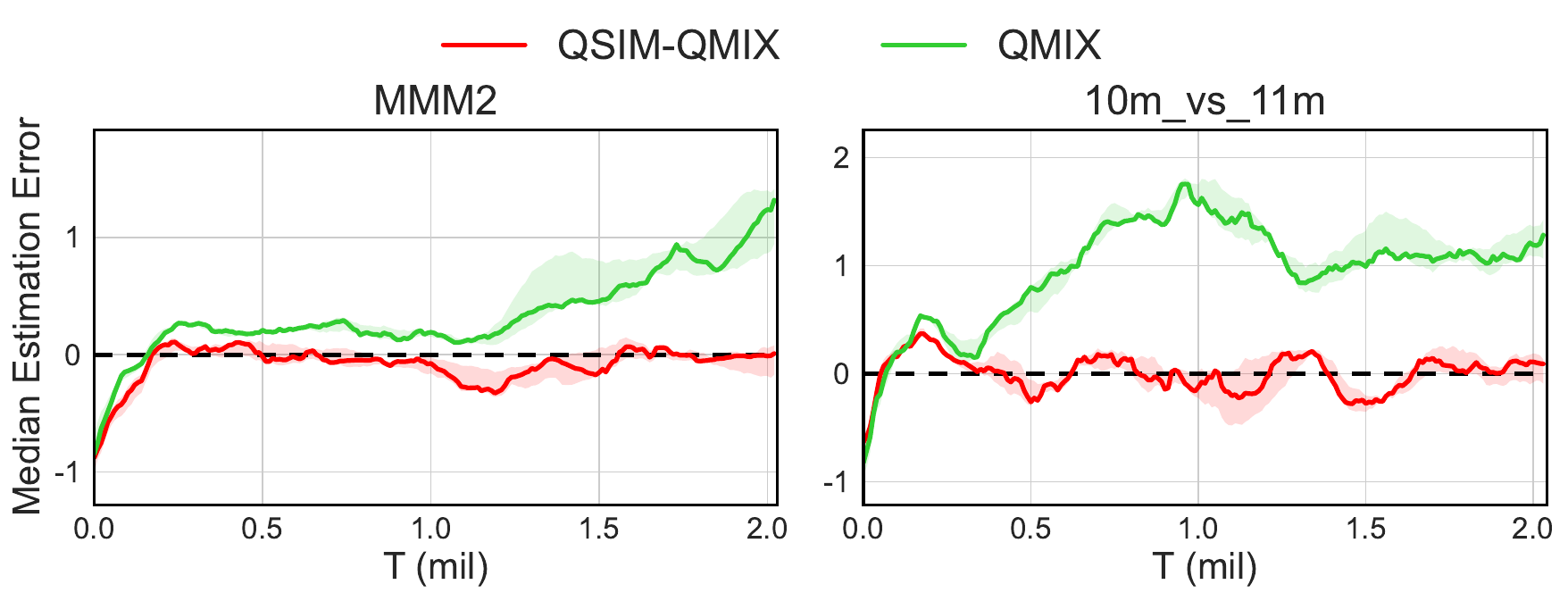}
    \caption{Comparison of Q-value estimation error $\delta_q$ on SMAC maps.}
    \label{fig:q_error}
    \vspace{-5pt}
\end{figure}

As illustrated in Figure~\ref{fig:q_error}, QSIM consistently exhibits lower estimation error $\delta_q$ compared with QMIX. 
The baseline QMIX suffers from a substantial accumulation of positive bias, stemming directly from its reliance on the standard max operator. 
In contrast, QSIM mitigates this bias by substituting the greedy target with a similarity weighted target. 
Consequently, the learned value estimates align more closely with the true returns, demonstrating that QSIM effectively mitigates overestimation and provides more accurate value estimates and a stable critic for VD methods.

\subsection{Visualization of Learned Action Representations}
\label{sec:exp_viz}

To assess whether the QSIM's feature encoder \(E_{\phi}\) learns action embeddings that capture meaningful action semantics, we visualize the representation space for a Stalker agent in SMAC \textit{3s5z} scenario, as shown in Figure~\ref{fig:embedding_viz}.
The t-SNE projection~\cite{t-SNE} in Figure~\ref{fig:embedding_viz}(b) reveals that actions with similar functional effects form coherent clusters.
For instance, ``Attack'' actions are grouped tightly, distinct from ``Move'' and ``No-op''. 
The similarity heatmap in Figure~\ref{fig:embedding_viz}(c) further supports this structure, showing high similarity scores among functionally related actions, with ``Attack’’ actions displaying strong pairwise affinity highlighted in red.
These results confirm that the learned action embeddings provide a reliable and interpretable metric for our similarity-based weighting scheme.

\section{Conclusion}

In this paper, we proposed QSIM to mitigate the systematic Q-value overestimation in VD methods. 
By replacing the greedy TD target in Bellman optimality equation with a similarity weighted TD target aggregated over a constructed near-greedy joint action space, QSIM effectively mitigates the maximization bias inherent in Q-learning. 
Extensive empirical evaluations on SMAC, SMACv2, MPE, and Matrix Games demonstrate that QSIM serves as a generalized module compatible with various VD methods. 
The results confirm that QSIM consistently improves the performance and stability of existing algorithms while significantly reducing the Q-value estimation error.

For future work, we plan to extend the paradigm of similarity weighted expectation to a broader range of multi-agent domains. 
Promising directions include integrating the QSIM mechanism into multi-agent actor-critic algorithms or adapting it for offline reinforcement learning settings, where robust value estimation is equally critical for learning stable policies from fixed datasets.

\section{Acknowledgments}
This work was supported by the National Natural Science Foundation of China (Grant No. 62506210).

\bigskip

\bibliography{aaai2026}

\newpage
\onecolumn
\appendix

\setcounter{secnumdepth}{3} 
\renewcommand{\thesection}{\Alph{section}} 

\section{\raggedright Further Details and Analysis of QSIM}
\label{app:qsim_details}

\subsection{Pseudocode for QSIM}
\label{app:pseudocode}
The complete training procedure for QSIM integrated with a generic value decomposition method is detailed in Algorithm~\ref{alg:qsim}.

\begin{algorithm}[htbp]
\caption{QSIM: Weighted Q-learning with Action Similarity}
\label{alg:qsim}
\begin{algorithmic}[1]
    \STATE \textbf{Initialize} network parameters $\theta$, target parameters $\theta^-$, encoder parameters $\phi$, learning rates $\alpha_\phi, \alpha_\theta$, replay buffer $\mathcal{D} = \{\}$.
    \WHILE{$step < step_{max}$}
        \STATE Collect experience and store in $\mathcal{D}$.
        \IF{$|\mathcal{D}| > \text{batch-size}$}
            \STATE Sample a batch of transitions $\{ (s, \bm{u}, r, s', \bm{u'}, \dots) \}$ from $\mathcal{D}$.
            
            \STATE Update the autoencoder parameters $\phi$ by minimizing the reconstruction loss $\mathcal{L}_{\text{AE}}$ according to Eq.~\eqref{eq:ae_loss}.
            \STATE Identify the greedy joint action $\bm{u}^*$ at the next state $s'$ using the main network.
            \STATE Generate action embeddings for greedy and deviating actions via the feature encoder $E_\phi$ using Eq.~\eqref{eq:embedding}.
            \STATE Calculate the cosine similarity scores between each deviating action and the greedy action using Eq.~\eqref{eq:similarity}.
            \STATE Derive the normalized weights $w(\bm{c})$ using the softmax function with inverse temperature $\kappa$ using Eq.~\eqref{eq:softmax_weights}.
            \STATE Construct the near-greedy joint action space $\mathcal{C}$ based on Eq.~\eqref{eq:suboptimal_space_def}.
            \STATE Calculate the candidate TD target $y(\bm{c})$ for each near-greedy joint action $\bm{c} \in \mathcal{C}$ using Eq.~\eqref{eq:individual_td_target}.
            \STATE Compute the final similarity weighted TD target $Y_{\text{QSIM}}$ as formulated in Eq.~\eqref{eq:final_target}.
            \STATE Update the network parameters $\theta$ by minimizing the TD error between $Y_{\text{QSIM}}$ and $Q_{\text{tot}}(\bm{\tau}, \bm{u}; \theta)$ defined in Eq.~\eqref{eq:td_loss}.
        \ENDIF
        
        \IF{$step$ \% update\_interval $= 0$}
            \STATE Update target network parameters: $\theta^- \leftarrow \theta$.
        \ENDIF
    \ENDWHILE
\end{algorithmic}
\end{algorithm}

\subsection{Autoencoder Training Convergence}
\label{app:ae_convergence}

In this section, we examine the training stability of the action representation learning module. We track the evolution of the next-observation prediction loss $\mathcal{L}_{\text{AE}}$ as defined in Eq.~\eqref{eq:ae_loss} across three diverse benchmark domains: SMAC, MPE, and Matrix Games. 

\begin{figure}[h]
    \centering
    \includegraphics[width=\linewidth]{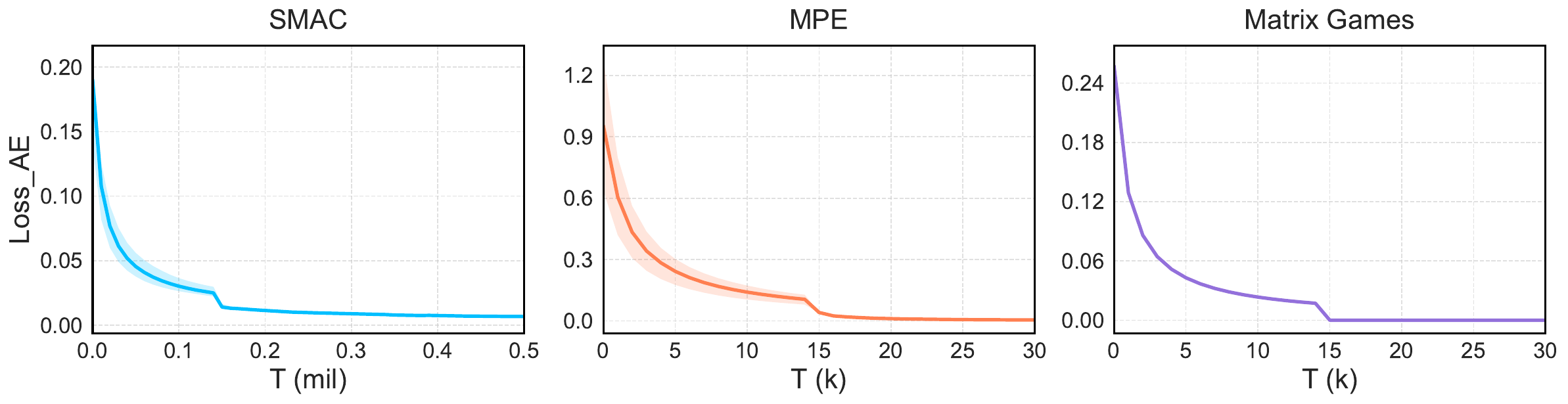}
    \caption{The evolution of autoencoder loss $\mathcal{L}_{\text{AE}}$ across SMAC, MPE, and Matrix Games scenarios. The consistent rapid drop in the early stages indicates efficient learning of environmental dynamics across diverse task complexities.}
    \label{fig:Loss_AE}
\end{figure}

As illustrated in Figure~\ref{fig:Loss_AE}, we observe a consistent trend across all scenarios: the reconstruction loss decreases sharply and stabilizes within the very early stages of training. This universal fast convergence indicates that the auxiliary task is sample-efficient and robust to varying task complexities. Crucially, this rapid stabilization ensures that the generated state-action embeddings become informative almost immediately. Consequently, the similarity weighted aggregation mechanism receives high-quality guidance from the outset of the policy learning phase, rather than being misled by unstable or random features.

\subsection{Additional Details on the Near-greedy Action Space}
\label{app:action_space}

The construction of the near-greedy joint action space $\mathcal{C}$, is a cornerstone of the QSIM framework. While a straightforward adaptation of Expected SARSA would involve computing an expectation over the entire joint action space $\mathcal{U}$, such an approach encounters two prohibitive obstacles in the multi-agent domain.

First, the Computational Intractability: The size of $\mathcal{U}$ scales exponentially with the number of agents ($|\mathcal{U}| = |\mathcal{A}|^N$), rendering full expectation computation infeasible for all but the simplest scenarios.
Second, the Estimation Unreliability: As highlighted in prior work~\cite{RES}, in high-dimensional spaces, the Q-values $Q_{tot}(s, \bm{u})$ for the majority of joint actions are rarely visited and thus poorly estimated. Naively incorporating these unreliable estimates into the target calculation introduces significant noise, potentially corrupting the learning signal.

\begin{figure}[h!]
    \centering
    \includegraphics[width=\linewidth]{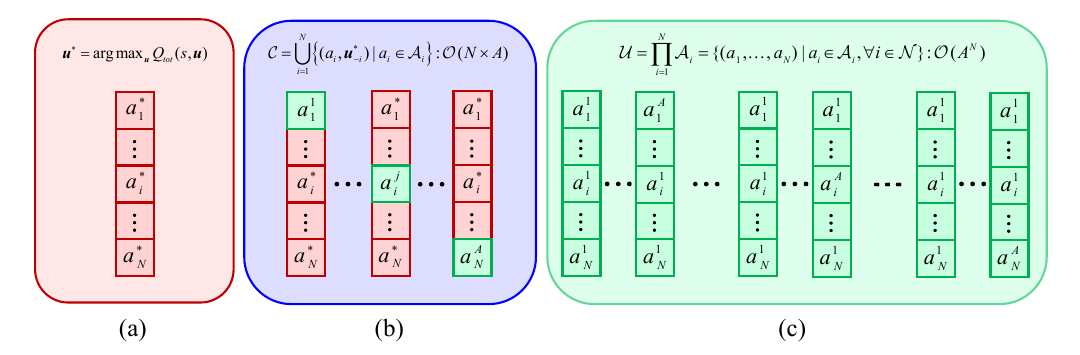}
    \caption{Schematic visualization of the action space construction. (a) A single greedy joint action $\bm{u}^*$. (b) The proposed near-greedy joint action space $\mathcal{C}$, constructed via single-agent deviations around the greedy anchor. (c) The exponentially large, intractable full joint action space $\mathcal{U}$.}
    \label{fig:suboptimal_space}
\end{figure}

To address these challenges, we construct a tractable subspace $\mathcal{C}$ centered around the greedy joint action $\bm{u}^*$, as visualized in Figure~\ref{fig:suboptimal_space}. Leveraging the IGM principle inherent in value decomposition, the greedy joint action $\bm{u}^*$ serves as a high-quality ``anchor'' in the action space. By limiting $\mathcal{C}$ to the union of all single-agent deviations from $\bm{u}^*$, we ensure that the included joint actions remain in the local neighborhood of the optimal policy. These actions are topologically close to the anchor and are significantly more likely to possess reliable value estimates compared to arbitrary samples from the vast full space $\mathcal{U}$.

This construction offers an optimal trade-off between efficiency and stability. It reduces the complexity from exponential to linear ($|\mathcal{C}| = N \times |\mathcal{A}|$), ensuring scalability. Simultaneously, it provides a robust basis for the weighted TD target by focusing the expectation on the most trustworthy region of the value function.

\subsection{Bias Analysis of QSIM}
\label{app:bias_analysis}

A primary motivation for QSIM is to mitigate the overestimation bias inherent in the construction of the TD target. Here, we provide a detailed analysis of the value estimation component of the TD target, demonstrating that the QSIM operator is inherently less prone to overestimation than the standard greedy max operator.

Let \(Q_{\text{tar}}(\bm{\tau}', \cdot; \theta^-)\) denote the joint action-values for the next joint history \(\bm{\tau}'\) produced by the target network with parameters \(\theta^-\). The standard TD target relies on a greedy value estimate. We denote this greedy estimate as \(V_{\text{Greedy}}(s')\):
\begin{equation*}
    V_{\text{Greedy}}(s') = \max_{\bm{u}'} Q_{\text{tar}}(\bm{\tau}', \bm{u}'; \theta^-) = Q_{\text{tar}}(\bm{\tau}', \bm{u}^*; \theta^-),
\end{equation*}
where \(\bm{u}^*\) is the greedy joint action defined in Eq.~\eqref{eq:greedy_action_selection}. The core of the overestimation bias stems from the fact that this max operator is applied over noisy Q-value estimates.

In contrast, for QSIM, the TD target is \(y_{\text{QSIM}} = r + \gamma V_{\text{QSIM}}(s')\). The value estimation component \(V_{\text{QSIM}}(s')\) is a similarity weighted expectation over the near-greedy action space \(\mathcal{C}\):
\begin{equation*}
    V_{\text{QSIM}}(s') = \sum_{\bm{c} \in \mathcal{C}} w(\bm{c}) Q_{\text{tar}}(\bm{\tau}', \bm{c}; \theta^-).
\end{equation*}

\noindent\textbf{Theorem 2.} \textit{The value estimation component of the QSIM TD target, denoted as \(V_{\text{QSIM}}(s')\), constitutes a lower bound on the standard greedy value estimate \(V_{\text{Greedy}}(s')\). Formally, since \(\sum_{\bm{c} \in \mathcal{C}} w(\bm{c}) = 1\) and \(w(\bm{c}) \ge 0\), it holds that:}
\begin{equation*}
    V_{\text{QSIM}}(s') \leq V_{\text{Greedy}}(s').
\end{equation*}

\noindent \textit{Proof.} By definition, \(V_{\text{QSIM}}(s')\) is a convex combination of the target Q-values within the near-greedy space \(\mathcal{C}\), as the weights \(w(\bm{c})\) are derived from a softmax function and satisfy \(w(\bm{c}) \ge 0\) and \(\sum_{\bm{c} \in \mathcal{C}} w(\bm{c}) = 1\).

For any specific action \(\bm{c} \in \mathcal{C}\), the inequality \(Q_{\text{tar}}(\bm{\tau}', \bm{c}; \theta^-) \leq \max_{\bm{c} \in \mathcal{C}} Q_{\text{tar}}(\bm{\tau}', \bm{c}; \theta^-)\) naturally holds. We can then expand the expectation term:
\begin{align*}
    V_{\text{QSIM}}(s') &= \sum_{\bm{c} \in \mathcal{C}} w(\bm{c}) Q_{\text{tar}}(\bm{\tau}', \bm{c}; \theta^-) \nonumber \\
    &\leq \sum_{\bm{c} \in \mathcal{C}} w(\bm{c}) \left( \max_{\bm{c} \in \mathcal{C}} Q_{\text{tar}}(\bm{\tau}', \bm{c}; \theta^-) \right) \nonumber \\
    &= \left( \max_{\bm{c} \in \mathcal{C}} Q_{\text{tar}}(\bm{\tau}', \bm{c}; \theta^-) \right) \sum_{\bm{c} \in \mathcal{C}} w(\bm{c}) \nonumber \\
    &= \max_{\bm{c} \in \mathcal{C}} Q_{\text{tar}}(\bm{\tau}', \bm{c}; \theta^-).
\end{align*}
Next, we recall the construction of the near-greedy joint action space \(\mathcal{C}\) defined in Eq.~\eqref{eq:suboptimal_space_def}. The set \(\mathcal{C}\) is formed by single-agent deviations from the greedy joint action \(\bm{u}^*\). This definition inherently implies that the greedy joint action itself is included in the set (i.e., the case where the deviating agent simply selects its greedy action, or effectively, the union covers the anchor point). Therefore, \(\bm{u}^* \in \mathcal{C}\).

Since \(\bm{u}^*\) is defined as the global maximizer (or the approximation thereof found by the mixing network), the maximum value within the subset \(\mathcal{C}\) is exactly the value of the greedy action:
\begin{equation*}
    \max_{\bm{c} \in \mathcal{C}} Q_{\text{tar}}(\bm{\tau}', \bm{c}; \theta^-) = Q_{\text{tar}}(\bm{\tau}', \bm{u}^*; \theta^-) = V_{\text{Greedy}}(s').
\end{equation*}
Substituting this back into the inequality, we conclude:
\begin{equation*}
    V_{\text{QSIM}}(s') \leq V_{\text{Greedy}}(s').
\end{equation*}

This theoretical result provides the basis for QSIM's stability. In deep Q-learning, value estimates \(Q_{\text{tar}}\) are inevitably corrupted by function approximation errors. The \(\max\) operator used in \(V_{\text{Greedy}}\) is highly sensitive to this noise, often selecting an action whose value is overestimated. This leads to \(V_{\text{Greedy}}(s')\) being a systematically biased estimator.
By replacing the hard max with a weighted average, QSIM produces a more conservative estimate that smooths out the noise from any single action's Q-value. This inherent dampening of noise spikes directly counteracts the maximization bias at its source, leading to more stable and accurate TD targets.

\subsection{Computational Efficiency and Runtime Analysis}
\label{app:runtime}

We evaluated the runtime impact of the extra components in QSIM, specifically the action-representation autoencoder. Despite the addition of this module, the computational cost remains minimal due to its lightweight MLP-based architecture, with each MLP consisting of only two linear layers and a ReLU activation function. Benchmarking results on the SMAC environment show that QSIM-QMIX's average training time (231.964ms/step) is nearly indistinguishable from that of QMIX (231.186ms/step). These results demonstrate that QSIM introduces negligible runtime overhead while significantly improving learning stability and performance, proving its practical viability for large-scale MARL tasks.

\section{\raggedright Experiment Details}
\label{app:experiment_details}

\subsection{Hyperparameters}
\label{app:hyperparameters}

We implement QSIM within the widely adopted PyMARL ecosystem. Specifically, the experiments on SMAC are conducted using the original PyMARL framework~\cite{SMAC}, while the evaluations for SMACv2, MPE and Matrix Games are implemented based on EPyMARL~\cite{epymarl}. Table~\ref{tab:common_hyperparameters} summarizes the common hyperparameters used for both the baseline algorithms and QSIM across these four benchmark environments. These settings are selected to match standard configurations in the literature, ensuring a fair comparison. Consistent with standard practice, all algorithms utilize an $\epsilon$-greedy exploration strategy, where $\epsilon$ is linearly annealed from $1.0$ to $0.05$ over the duration defined by the Epsilon Anneal Steps. Finally, Table~\ref{tab:qsim_hyperparameters} lists the specific hyperparameters introduced by the QSIM framework.

\begin{table}[h!]
\centering
\renewcommand{\arraystretch}{1.2}
\begin{tabular}{l|cccc}
\toprule
\textbf{Parameter} & \textbf{SMAC} & \textbf{SMACv2} & \textbf{MPE} & \textbf{Matrix Games} \\
\midrule
Network Type & GRU & GRU & GRU & FC \\
Learning Rate & 0.0005 & 0.0005 & 0.0005 & 0.0005 \\
Buffer Size & 5000 & 5000 & 5000 & 5000 \\
Epsilon Anneal Steps & 50,000 & 50,000 & 200,000 & 50,000 \\
Target Update Interval & 200 (Hard) & 200 (Hard) & 200 (Hard) & 0.01 (Soft) \\
Reward Standardization & False & True & True & True \\
Discount Factor & 0.99 & 0.99 & 0.99 & 0.99 \\
\bottomrule
\end{tabular}
\caption{Common hyperparameters used in different environments.}
\label{tab:common_hyperparameters}
\end{table}

\begin{table}[h!]
\centering
\renewcommand{\arraystretch}{1.2}
\begin{tabular}{l|c}
\toprule
\textbf{Parameter} & \textbf{Value} \\
\midrule
Autoencoder Hidden Dimension & 128 \\
Inverse Temperature $\kappa$ & 3 \\
Similarity Threshold & 0 \\
\bottomrule
\end{tabular}
\caption{QSIM-specific hyperparameters.}
\label{tab:qsim_hyperparameters}
\end{table}

Regarding the \textbf{Similarity Threshold}, we set this value to 0 to refine the weighting process. After computing the pairwise cosine similarity matrix $S$, actions with a similarity score $S_{ij} < 0$ are considered functionally dissimilar or contradictory to the greedy action. We believe these actions provide little to no constructive information for value estimation. Therefore, they are masked before the softmax normalization, ensuring that the weighted TD target calculation is focused solely on the relevant subspace of actions that share positive semantic alignment with the optimal policy.

\subsection{Hardware and Software Infrastructure}
\label{app:infrastructure}

All experiments were conducted on a high-performance computing cluster to ensure efficiency and reproducibility. The specific hardware and software configurations are detailed as follows:

\begin{itemize}
    \item \textbf{Hardware Setup:} The experiments were executed on computing nodes equipped with NVIDIA H100 PCIe GPUs (80GB VRAM) and an Intel(R) Xeon(R) Platinum 8488C CPU.
    
    \item \textbf{Software Environment:} Our framework is implemented in Python 3.8.20. We utilize PyTorch 2.1.0 as the deep learning backend. The system environment is configured with NVIDIA Driver 570.195.03 and CUDA 12.8 to support the computational requirements.
\end{itemize}

\section{\raggedright Additional Results}
\label{app:add_results}

\subsection{Performance on SMACv2}
\label{app:smacv2}

To further evaluate the robustness of QSIM in highly stochastic and non-stationary environments, we extend our experiments to the StarCraft Multi-Agent Challenge v2 (\textbf{SMACv2}). SMACv2 significantly enhances environmental stochasticity through procedural content generation, specifically by randomizing unit start locations and unit types in each episode. This dynamic setting necessitates that agents learn generalized policies capable of adapting to unseen configurations, providing a rigorous testbed for assessing the algorithm's adaptability and stability under uncertainty.

\begin{figure}[h!]
    \centering
    \includegraphics[width=\linewidth]{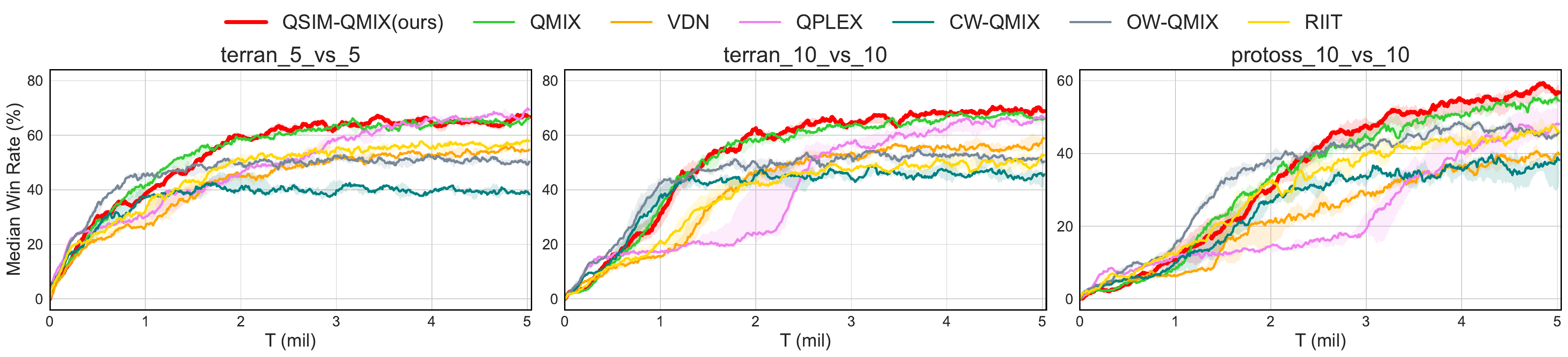}
    \caption{Performance comparison on SMACv2 maps.}
    \label{fig:SMACv2}
\end{figure}

We evaluate QSIM-QMIX against baselines on three representative scenarios: \textit{terran\_5\_vs\_5}, \textit{terran\_10\_vs\_10} and \textit{protoss\_10\_vs\_10}. The comparative results are visualized in Figure~\ref{fig:SMACv2}.

As observed, QSIM-QMIX continues to outperform the baseline algorithms, demonstrating its versatility. However, we note that the performance margin over QMIX is narrower compared to the original SMAC results. We attribute this phenomenon to the inherent stochasticity of SMACv2. In QSIM, the effectiveness of the similarity weighting relies on the autoencoder learning stable and consistent semantic representations of actions. In SMACv2, the meaning or outcome of an action (e.g., ``Attack'') can vary drastically across episodes depending on the randomized unit types and positions. This variability introduces noise into the learned action embeddings, making the similarity metric less distinct than in deterministic environments. Nevertheless, despite this challenge, QSIM still provides a more robust learning signal than the standard greedy approach, confirming its efficacy even under conditions of high environmental randomness.

\subsection{Additional Generality Experiments}
\label{app:add_generality}

\begin{figure}[ht]
    \centering
    \includegraphics[width=\linewidth]{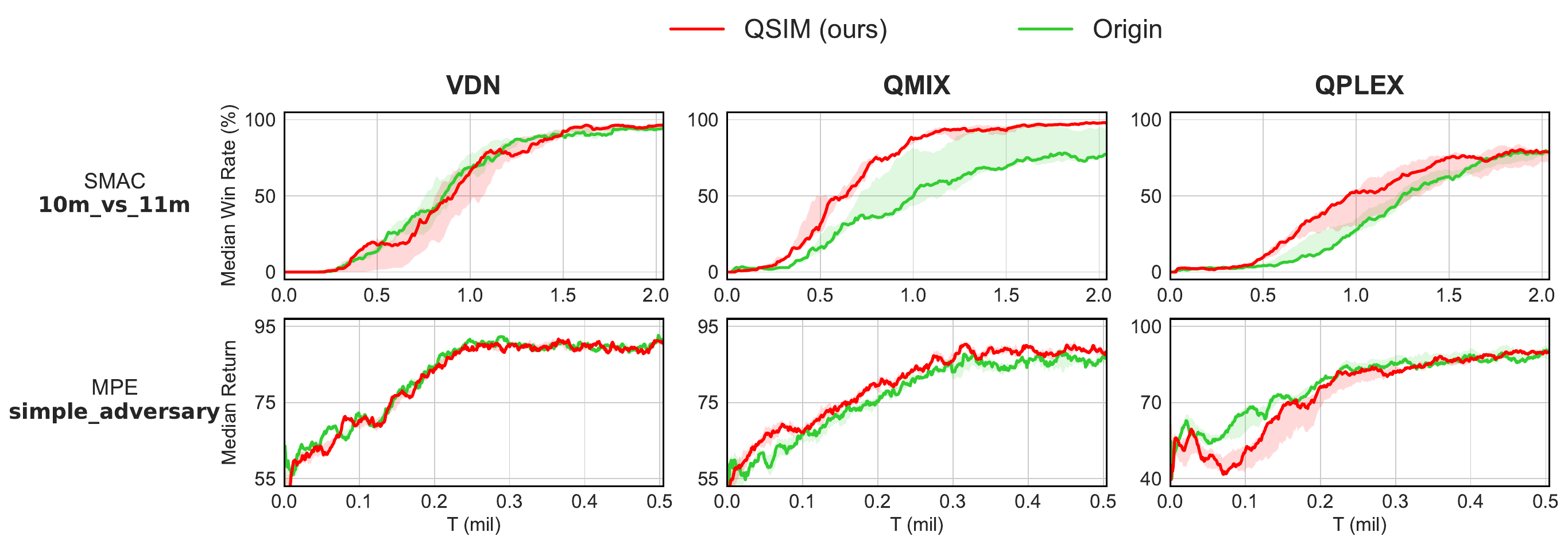} 
    \caption{Performance comparison of QSIM-enhanced algorithms on \textit{10m\_vs\_11m} and \textit{simple\_adversary}.}
    \label{fig:app_generality}
\end{figure}

Due to space constraints in the main text, we present the generality evaluation results for the remaining scenarios in this section. Figure~\ref{fig:app_generality} illustrates the performance of QSIM integrated with QMIX, VDN, and QPLEX on the SMAC map \textit{10m\_vs\_11m} and the MPE scenario \textit{simple\_adversary}.

Consistent with the findings in Section~\ref{sec:generality}, the QSIM-enhanced algorithms exhibit superior convergence speed and final performance compared to their vanilla counterparts. These results further corroborate the universal effectiveness of the QSIM module across varying environment types and backbone algorithms.

\subsection{Additional Ablation Study}
\label{app:add_abalation}

Beyond the core components analyzed in the main text, we investigate two additional structural variants to assess the impact of specific design choices in the QSIM framework:

\begin{itemize}
    \item \textbf{QSIM-NoState:} In this variant, we modify the input to the self-supervised autoencoder. Instead of feeding the global state $s$ alongside the local observation $o_i$ and action $a_i$, we remove the global state. The encoder solely relies on $(o_i, a_i)$ to predict the next local observation. This ablation tests whether local information is sufficient for learning meaningful action semantics.
    \item \textbf{QSIM-TopN:} In this variant, we alter the aggregation scheme for the weighted TD target. Instead of computing a weighted average over the entire near-greedy action space $\mathcal{C}$, we only select the top-$N$ actions with the highest similarity scores. The weights are then renormalized over this truncated subset. This allows us to investigate whether actions with lower functional similarity contribute constructively to the value update, or if the estimation benefits from strictly excluding them.
\end{itemize}

\begin{figure}[ht]
    \centering
    \includegraphics[width=0.5\linewidth]{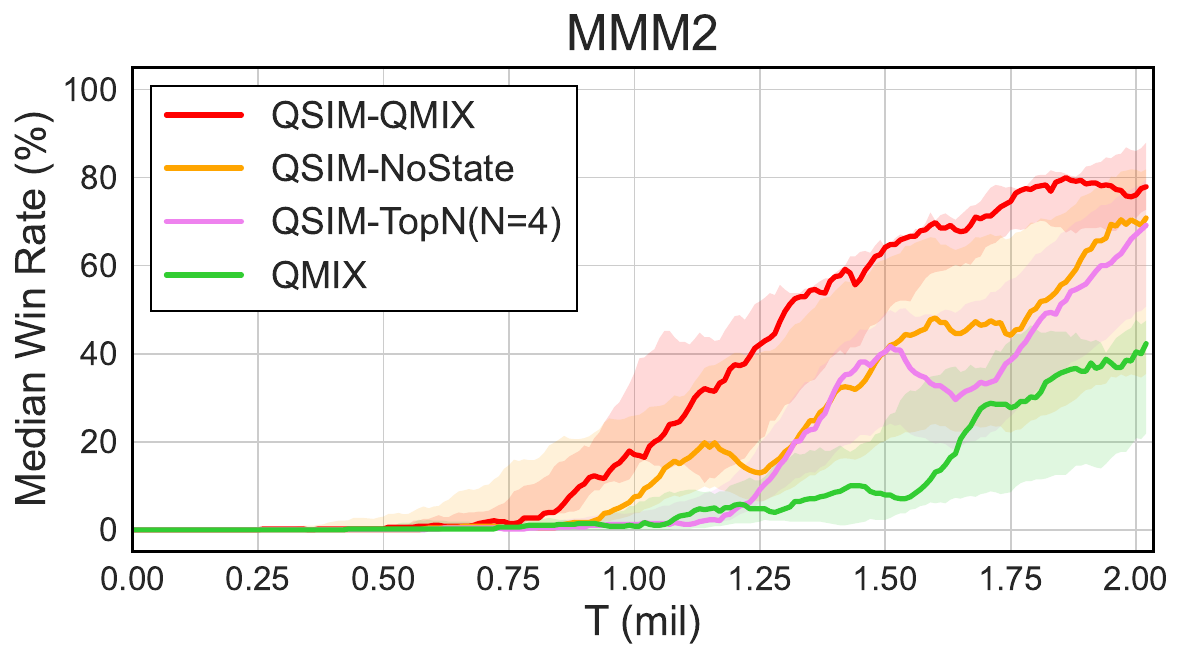}
    \caption{Ablation study on QSIM-NoState and QSIM-TopN.}
    \label{fig:add_ablation}
\end{figure}

The comparative results are presented in Figure~\ref{fig:add_ablation}. 
First, the distinct performance gap between QSIM and QSIM-NoState highlights the critical role of the global state $s$ in action representation learning. By incorporating the global view, the autoencoder can better capture complex environmental dynamics that are partially unobservable from a local perspective, thereby producing more accurate and semantically meaningful action embeddings.

Second, regarding the aggregation strategy, we observe that QSIM consistently outperforms QSIM-TopN. This finding is pivotal: it suggests that actions with lower similarity scores are not merely noise or irrelevant. Instead, they provide a necessary regularization effect that smooths the value landscape. Aggressively truncating these actions diminishes this smoothing benefit and re-introduces higher variance into the target calculation. Consequently, we adopt the strategy of aggregating over the entire near-greedy action space $\mathcal{C}$ to ensure the most robust and stable TD target construction.

\subsection{Sensitivity Analysis of $\kappa$}
\label{app:kappa_analtsis}

\begin{figure}[ht]
    \centering
    \includegraphics[width=0.5\linewidth]{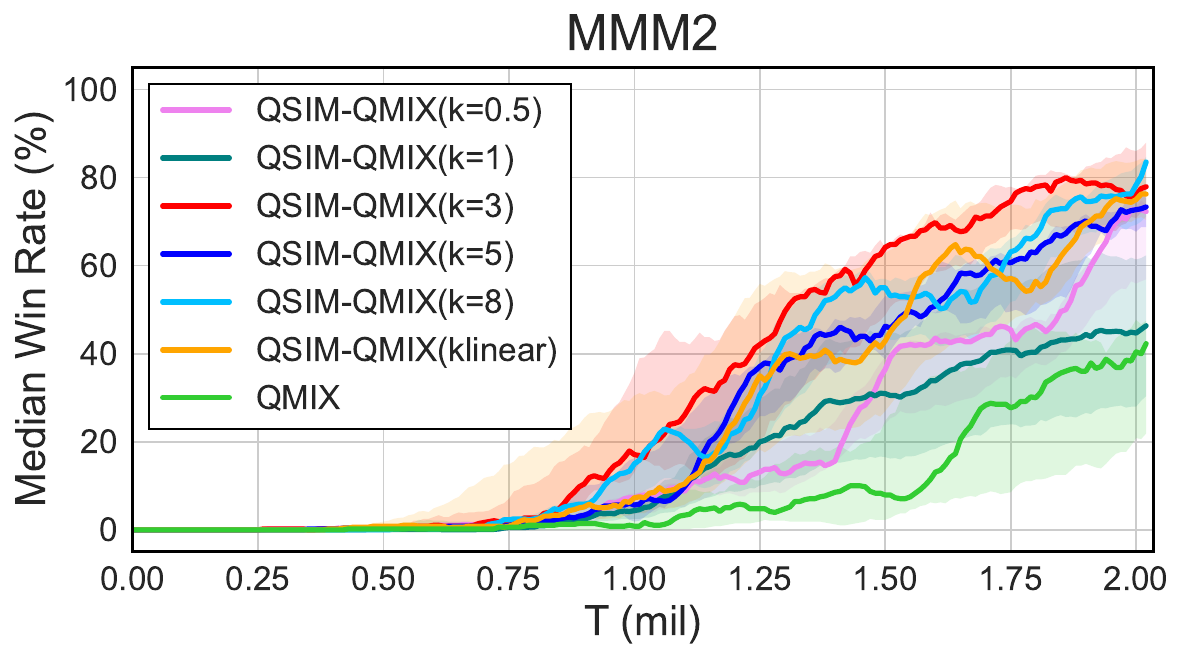}
    \caption{Sensitivity analysis of the inverse temperature parameter $\kappa$ on \textit{MMM2}.}
    \label{fig:kappa_sensitivity}
\end{figure}

The inverse temperature parameter $\kappa$ in Eq.~\eqref{eq:softmax_weights} governs the sharpness of the similarity-based weighting distribution in QSIM. A lower $\kappa$ approaches a uniform average, while a higher $\kappa$ approximates the greedy max operator. To identify the optimal configuration, we conduct a sensitivity analysis on SMAC maps comparing distinct fixed values (e.g., 0.5, 1, 3, 5, 8) against a dynamic scheduling variant, \textbf{QSIM-klinear}. The results are visualized in Figure~\ref{fig:kappa_sensitivity}.

In QSIM-klinear, $\kappa$ linearly increases from 1 to 10 over the course of training, designed to transition from a high-variance smoothing regime (prioritizing exploration) to a more aggressive optimality focus (prioritizing exploitation) as the policy matures. However, our empirical results indicate that while dynamic scheduling offers theoretical benefits, a carefully tuned fixed value provides more stable performance. Specifically, we observe that setting $\kappa=3$ consistently yields the superior results across the majority of scenarios, striking the most effective balance between mitigating overestimation and retaining policy precision. Consequently, we adopt $\kappa=3$ as the default hyperparameter for our main experiments.

\section{\raggedright Theoretical Analysis of Maximization Bias}
\label{app:bias_proof}

In this section, we provide the detailed proof for the scaling of the maximization bias upper bound under Gaussian noise assumptions, as presented in Theorem 1 of the main text.

\subsection{Problem Setup and Theorem}

Consider the joint action space $\mathcal{U}$ with size $M = |\mathcal{U}| = |\mathcal{A}|^N$. Let the estimated joint Q-value be modeled as $\hat{Q}(\bm{u}) = Q^*(\bm{u}) + \epsilon_{\bm{u}}$, where $\epsilon_{\bm{u}} \sim \mathcal{N}(0, \sigma^2)$ are Gaussian noise terms. To analyze the pure maximization bias, we consider the worst-case scenario where true values are identical: $Q^*(\bm{u}) = \mu$ for all $\bm{u}$.

\textbf{Theorem 1.} \textit{The maximization bias is upper bounded as follows:}
\begin{equation*}
    \mathbb{E}\left[\max_{\bm{u}} \hat{Q}(\bm{u})\right] - \max_{\bm{u}} Q^*(\bm{u}) \le \sigma \sqrt{2 \ln |\mathcal{A}|^N}.
\end{equation*}

\subsection{Proof}
\noindent \textit{Proof.} Let the standardized noise variables be $Z_{\bm{u}} \sim \mathcal{N}(0, 1)$, such that $\hat{Q}(\bm{u}) = \mu + \sigma Z_{\bm{u}}$. The bias is defined as:
\begin{equation*}
    \text{Bias} = \mathbb{E}\left[\max_{\bm{u}} \hat{Q}(\bm{u})\right] - \mu = \sigma \mathbb{E}\left[\max_{\bm{u}} Z_{\bm{u}}\right].
\end{equation*}
Let $Z_{\max} = \max_{\bm{u}} Z_{\bm{u}}$. To bound $\mathbb{E}[Z_{\max}]$, we introduce a scalar $t > 0$ and consider the exponential function $f(x) = e^{tx}$. Since $f(x)$ is a convex function, we can apply Jensen's inequality:
\begin{equation*}
    e^{t \mathbb{E}[Z_{\max}]} \leq \mathbb{E}\left[e^{t Z_{\max}}\right].
\end{equation*}
Next, we utilize the monotonic property of the exponential function. Since $e^{tx}$ is strictly increasing for $t > 0$, the exponential of the maximum value is equivalent to the maximum of the exponential values:
\begin{equation*}
    \mathbb{E}\left[e^{t Z_{\max}}\right] = \mathbb{E}\left[\max_{\bm{u}} e^{t Z_{\bm{u}}}\right].
\end{equation*}
Since the exponential term $e^{t Z_{\bm{u}}}$ is always positive, the maximum of these terms is strictly bounded by their sum (i.e., $\max_i x_i \leq \sum_i x_i$ for $x_i > 0$). Applying the linearity of expectation, we obtain:
\begin{equation*}
    \mathbb{E}\left[\max_{\bm{u}} e^{t Z_{\bm{u}}}\right] \leq \mathbb{E}\left[\sum_{\bm{u}=1}^M e^{t Z_{\bm{u}}}\right] = \sum_{\bm{u}=1}^M \mathbb{E}\left[e^{t Z_{\bm{u}}}\right].
\end{equation*}
Now, we recall the property of the Moment Generating Function (MGF) for a standard normal variable $Z \sim \mathcal{N}(0, 1)$, which states that $\mathbb{E}[e^{tZ}] = e^{t^2/2}$. Substituting this into the sum:
\begin{equation*}
    \sum_{\bm{u}=1}^M \mathbb{E}\left[e^{t Z_{\bm{u}}}\right] = \sum_{\bm{u}=1}^M e^{t^2/2} = M e^{t^2/2}.
\end{equation*}
Combining the steps above, we have the inequality:
\begin{equation*}
    e^{t \mathbb{E}[Z_{\max}]} \leq M e^{t^2/2}.
\end{equation*}
Taking the natural logarithm on both sides yields:
\begin{equation*}
    t \mathbb{E}[Z_{\max}] \leq \ln M + \frac{t^2}{2} \quad \implies \quad \mathbb{E}[Z_{\max}] \leq \frac{\ln M}{t} + \frac{t}{2}.
\end{equation*}
To find the tightest bound, we minimize the right-hand side with respect to $t$. Setting the derivative to zero yields the optimal $t^* = \sqrt{2 \ln M}$. Substituting $t^*$ back into the inequality:
\begin{equation*}
    \mathbb{E}[Z_{\max}] \leq \frac{\ln M}{\sqrt{2 \ln M}} + \frac{\sqrt{2 \ln M}}{2} = \sqrt{2 \ln M}.
\end{equation*}
Substituting this bound back into the bias equation* and recognizing that $M = |\mathcal{A}|^N$:
\begin{equation*}
    \text{Bias} = \sigma \mathbb{E}[Z_{\max}] \leq \sigma \sqrt{2 \ln M} = \sigma \sqrt{2 \ln (|\mathcal{A}|^N)}.
\end{equation*}

It is a well-established result in Extreme Value Theory that this upper bound is asymptotically tight for Gaussian random variables as $M \to \infty$. That is, the expected maximum converges to this bound: $\mathbb{E}[Z_{\max}] \approx \sqrt{2 \ln M}$. Given the exponential size of the joint action space in MARL, this bound accurately characterizes the worst-case scaling behavior of the bias. Consequently, we establish the upper bound relationship in Theorem 1:
\begin{equation*}
    \mathbb{E}\left[\max_{\bm{u}} \hat{Q}(\bm{u})\right] - \max_{\bm{u}} Q^*(\bm{u}) \le \sigma \sqrt{2 \ln |\mathcal{A}|^N}.
\end{equation*}

\section{\raggedright Introduction for Environments}
\label{app:env_intro}

In this section, we provide detailed descriptions of the three benchmark environments utilized in our experiments: the StarCraft Multi-Agent Challenge (SMAC), Multi-Agent Particle Environments (MPE), and Matrix Games.

\subsection{StarCraft Multi-Agent Challenge (SMAC)}
\label{app:smac_details}

\begin{figure}[h!]
    \centering
    \begin{subfigure}[b]{0.45\linewidth}
        \centering
        \includegraphics[width=6.5cm, height=4.5cm]{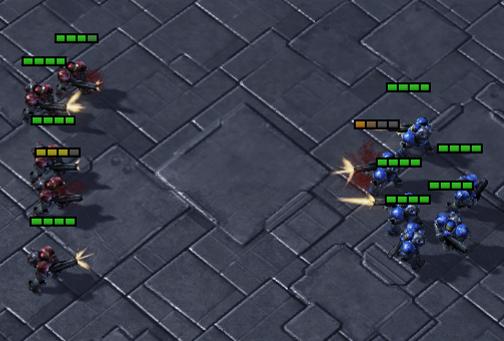}
        \caption{SMAC}
        \label{fig:smac_viz_a}
    \end{subfigure}
    \hspace{-0.5cm}
    \begin{subfigure}[b]{0.45\linewidth}
        \centering
        \includegraphics[width=6.5cm, height=4.5cm]{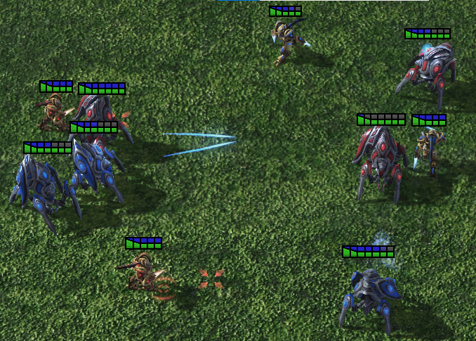}
        \caption{SMACv2}
        \label{fig:smac_viz_b}
    \end{subfigure}
    \caption{Visualization of scenarios in SMAC and SMACv2.}
    \label{fig:smac}
\end{figure}

SMAC is a high-dimensional micro-management benchmark based on the StarCraft II engine.A visual comparison of the scenarios in SMAC and SMACv2 is provided in Figure~\ref{fig:smac}.

For the SMAC benchmark, we select eight representative maps spanning three difficulty levels; the specific unit configurations for both allied and enemy teams are summarized in Table~\ref{tab:smac_challenges}. For the SMACv2 benchmark, our evaluation focuses on the Terran and Protoss races using the stochastic unit types and start position distribution. The specific generation probabilities for different unit types and distribution settings are detailed in Table~\ref{tab:smacv2_config}.

\begin{table}[h!]
\centering
\renewcommand{\arraystretch}{1.2}
{
\begin{tabular}{lccc}
\toprule
\textbf{Map Name} & \textbf{Ally Units} & \textbf{Enemy Units} & \textbf{Difficulty} \\
\midrule
\textit{2s3z} & 2 Stalkers \& 3 Zealots & 2 Stalkers \& 3 Zealots & Easy \\
\textit{1c3s5z} & 1 Colossus, 3 Stalkers \& 5 Zealots & 1 Colossus, 3 Stalkers \& 5 Zealots & Easy \\
\midrule
\textit{2c\_vs\_64zg} & 2 Colossi & 64 Zerglings & Hard \\
\textit{3s5z} & 3 Stalkers \& 5 Zealots & 3 Stalkers \& 5 Zealots & Hard \\
\textit{8m\_vs\_9m} & 8 Marines & 9 Marines & Hard \\
\textit{10m\_vs\_11m} & 10 Marines & 11 Marines & Hard \\
\midrule
\textit{MMM2} & 1 Medivac, 2 Marauders \& 7 Marines & 1 Medivac, 3 Marauders \& 8 Marines & Super-Hard \\
\textit{3s5z\_vs\_3s6z} & 3 Stalkers \& 5 Zealots & 3 Stalkers \& 6 Zealots & Super-Hard \\
\bottomrule
\end{tabular}%
}
\caption{Parameter configurations of SMAC scenarios.}
\label{tab:smac_challenges}
\end{table}

\begin{table}[h!]
\centering
\renewcommand{\arraystretch}{1.2}
\begin{tabular}{cccc}
\toprule
\textbf{Race} & \textbf{Unit Types} & \textbf{\makecell{Probability of \\ Unit Type Generation}} & \textbf{\makecell{Probability of \\ Start Position}} \\
\midrule
\multirow{3}{*}{Terran} & Marine & 0.45 & \multirow{3}{*}{\makecell{Surrounded: 0.5 \\ Reflect: 0.5}} \\
 & Marauder & 0.45 & \\
 & Medivac & 0.10 & \\
\midrule
\multirow{3}{*}{Protoss} & Stalker & 0.45 & \multirow{3}{*}{\makecell{Surrounded: 0.5 \\ Reflect: 0.5}} \\
 & Zealot & 0.45 & \\
 & Colossus & 0.10 & \\
\bottomrule
\end{tabular}
\caption{Parameter configurations of SMACv2 scenarios.}
\label{tab:smacv2_config}
\end{table}

\subsection{Multi-Agent Particle Environments (MPE)}
\label{app:mpe_details}

\begin{figure}[h!]
    \centering
    \begin{subfigure}[b]{0.4\linewidth}
        \centering
        \includegraphics[width=3.3cm, height=4.4cm]{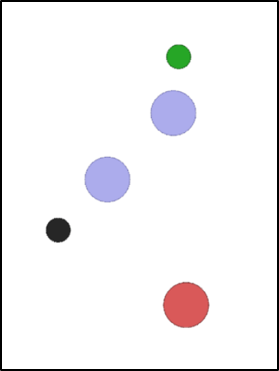}
        \caption{\textit{simple\_adversary}}
        \label{fig:mpe_adversary}
    \end{subfigure}
    \hspace{-3.5cm}
    \begin{subfigure}[b]{0.4\linewidth}
        \centering
        \includegraphics[width=3.3cm, height=4.4cm]{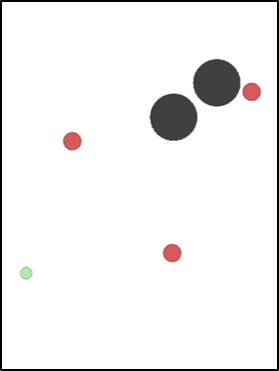}
        \caption{\textit{simple\_tag}}
        \label{fig:mpe_tag}
    \end{subfigure}
    \caption{Visualization of MPE scenarios.}
    \label{fig:mpe}
\end{figure}

In addition to the high-dimensional SMAC tasks, we utilize two scenarios from the Multi-Agent Particle Environments (MPE) to test mixed cooperative-competitive dynamics. The visualizations are shown in Figure~\ref{fig:mpe}. Details of the MPE environment parameters used in the experiments are shown in Table~\ref{tab:mpe_environments}.

\begin{itemize}
    \item \textbf{\textit{simple\_adversary}}: This environment consists of 1 adversary (red), 2 cooperative agents (blue), and 2 landmarks (black and green). While all agents can observe the positions of others and landmarks, only the cooperative agents know which landmark is the true target (green). The cooperative team is rewarded based on the proximity of their closest agent to the target, but penalized based on the adversary's distance to it. This dynamic necessitates complex coordination: the blue agents must learn to split up and cover both landmarks to effectively deceive the red adversary.

    \item \textbf{\textit{simple\_tag}}: This scenario represents a predator-prey dynamic. The adversary (green) is faster and incurs a negative reward upon collision with the agents (red). The slower agents, in turn, are rewarded for colliding with the adversary. Large black circles serve as obstacles that block movement. This scenario forces the slower agents to coordinate their actions to corner and trap the faster prey.
\end{itemize}

\begin{table}[h!]
\centering
\renewcommand{\arraystretch}{1.2}
{
\begin{tabular}{lcccccc}
\toprule
\textbf{Scenario} & \textbf{Agents} & \textbf{Action dim} & \textbf{State dim} & \textbf{Observation dim} \\
\midrule
\textit{simple\_adversary} & 2 Agent \& 1 Adversary & 5 & 28 & 10 \& 8 \\
\textit{simple\_tag} & 3 Agent \& 1 Adversary & 5 & 62 & 14 \& 16 \\
\bottomrule
\end{tabular}%
}
\caption{Parameter configurations of MPE scenarios.}
\label{tab:mpe_environments}
\end{table}

\subsection{Matrix Games}
\label{app:matrix_game_details}

We employ the classic \textit{climbing} scenario from Matrix Games to evaluate algorithm convergence speed and the ability to escape local optima. The payoff matrix is presented in Table~\ref{tab:matrix_game}.

This game poses a significant challenge for coordination. The agents must conceptually ``climb'' from the local optimum at joint action $(A, C)$ or $(B, B)$ towards the global optimum at $(C, A)$. Specifically, in this matrix formulation, the global optimal reward is 11 at $(C, A)$, while a local optimum exists at $(B, B)$ with a reward of 7. The difficulty arises from the severe penalty of -30. If the agents coordinate at $(B, B)$, a single deviation by either agent leads to this catastrophic penalty. Consequently, many algorithms fail to stabilize at the optimal solution or get stuck at safer, near-greedy equilibria to avoid the risk.

\begin{table}[h!]
    \centering
    \renewcommand{\arraystretch}{1.4} 
    \begin{tabular}{|p{1.5cm}<{\centering}||p{1.5cm}<{\centering}|p{1.5cm}<{\centering}|p{1.5cm}<{\centering}|}
        \hline
        \multicolumn{1}{|c||}{\diagbox[width=\dimexpr 1.5cm+2\tabcolsep\relax]{$n_1$}{$n_2$}} & A & B & C \\ 
        \hline \hline
        A & 0 & 6 & 5 \\
        \hline
        B & -30 & 7 & 0 \\
        \hline
        C & 11 & -30 & 0 \\
        \hline
    \end{tabular}
    \caption{Payoff matrix for the challenging \textit{climbing} game.}
    \label{tab:matrix_game}
\end{table}

\end{document}